\documentclass[draft]{agujournal2019}
\usepackage{url} 
\usepackage{lineno}
\usepackage{soul}
\usepackage{hyperref}
\usepackage{natbib}
\usepackage{amsmath}
\usepackage{amssymb}

\newcommand{\sti}[1]{\unskip}
\newcommand{\newTxt}{\textcolor{black}}

\renewcommand{\st}[1]{\unskip}
\newcommand{\newTxtR}{\textcolor{black}}

\draftfalse

\begin{document}

\title{Granger causal inference for climate change attribution}

\authors{Mark D. Risser\affil{1}, Mohammed Ombadi\affil{2}, Michael F. Wehner\affil{3}}

\affiliation{1}{Climate and Ecosystem Sciences Division, LBNL, Berkeley, CA, 94720.}
\affiliation{2}{College of Engineering, University of Michigan, Ann Arbor, MI, 48109.} 
\affiliation{3}{Applied Mathematics and Computational Research Division, LBNL, Berkeley, CA, 94720.}

\correspondingauthor{Mark D. Risser}{mdrisser@lbl.gov}
\vskip4ex 

\begin{abstract}
Climate change detection and attribution (D\&A) is concerned with determining the extent to which anthropogenic activities have influenced specific aspects of the global climate system. D\&A fits within the broader field of causal inference, the collection of statistical methods that identify cause and effect relationships. There are a wide variety of methods for making attribution statements, each of which require different types of input data and focus on different types of weather and climate events  and each of which are conditional to varying extents. Some methods are based on Pearl causality (direct experimental interference) while others leverage Granger (predictive) causality, and the causal framing provides important context for how the resulting attribution conclusion should be interpreted. However, while Granger-causal attribution analyses have become more common, there is no clear statement of their strengths and weaknesses relative to Pearl-causal attribution and no clear consensus on where and when Granger-causal perspectives are appropriate. In this prospective paper, we provide a formal definition for Granger-based approaches to trend and event attribution and a clear comparison with more traditional methods for assessing the human influence on extreme weather and climate events. Broadly speaking, Granger-causal attribution statements can be constructed quickly from observations and do not require computationally-intesive dynamical experiments. These analyses also enable rapid attribution, which is useful in the aftermath of a severe weather event, and provide multiple lines of evidence for anthropogenic climate change when paired with Pearl-causal attribution. Confidence in attribution statements is increased when different methodologies arrive at similar conclusions.  Moving forward, we encourage the D\&A community to embrace hybrid approaches to climate change attribution that leverage the strengths of both Granger and Pearl causality.

\begin{center}
\end{center}

\end{abstract}

\section{Introduction} \label{sec:intro}

Detection and attribution (D\&A) of anthropogenic climate change seeks to make cause and effect statements regarding if, how, and why different aspects of the global climate system have changed due to specific human activities. As the name suggests, D\&A is a two part exercise: first, for a given weather or climate variable of interest, one attempts to detect a systematic change in its distribution; and second, if a change can be detected, one further tries to attribute or ascribe the identified changes to specific external forcing agents. 
Leveraging methods developed by \cite{hasselmann1979signal,Hasselmann1993optimal}, \cite{North1995}, and \cite{Allen2003estimating}, there are now many decades of literature on important D\&A results documenting the human influence on, among other things, surface air temperature \citep{hegerl1997multi,tett1999causes}, sea level pressure \citep{gillett2003detection}, tropopause height \citep{santer2003contributions}, free atmospheric temperature \citep{jones2003causes}, ocean heat content \citep{barnett2005penetration}, and precipitation \citep{zhang2007detection}. While the earliest results involved D\&A for large-scale averages on long time  scales, robust attribution conclusions have more recently been made for extremes \citep{Min2011human,Kim2015attribution}, individual weather and climate events \citep{Pall2017diagnosing,Risser2017,Patricola2018anthropogenic,van2019human}, and even the impacts of individual extreme events \citep{Frame2020,Wehner2021attributable,Smiley2022social, Mitchell_2016_mortality, Kirchmeier}. Attribution conclusions are critical for climate change adaptation and mitigation \citep{IPCC_AR6_WG3}, improving future climate projections \citep{Mitchell2017half}, and even determining the losses and damages associated with human-induced climate change \citep{Noy2023event}.

\begin{figure}[!t]
    \centering
    \includegraphics[trim={155 25 85 30mm}, clip, width=0.8\textwidth]{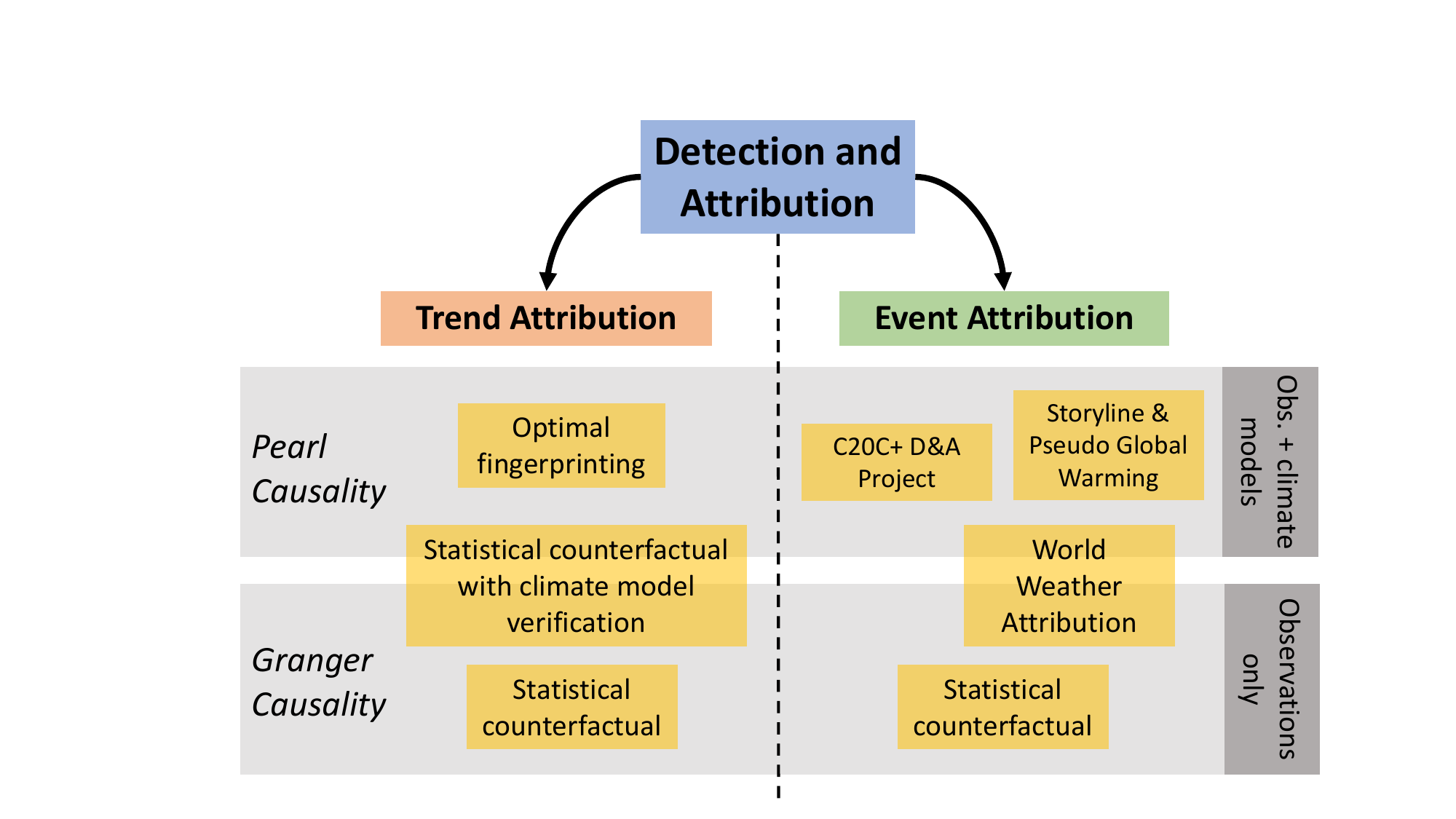}
    \caption{A schematic overview of the commonly-used approaches to climate change Detection \& Attribution, considering both trend and event attribution methods. Each method is categorized according to whether it utilizes Pearl or Granger causality.}
    \label{fig_overview}
\end{figure}

Climate change attribution studies can be broadly placed into two categories: \textit{trend attribution} and \textit{event attribution}; see Figure~\ref{fig_overview}. Trend attribution methods evaluate long-term changes to the global climate system, typically considering climatological changes over (at least) several decades. Nearly all early trend attribution studies use optimal fingerprinting \citep[][]{Hasselmann1993optimal, Allen2003estimating}, which uses a statistical linear regression model to project observed patterns of change onto corresponding model-simulated patterns of change. The optimal fingerprinting method remains widely used, even in the most recent attribution studies \citep{christidis2022human}. The attribution conclusion in optimal fingerprinting is determined by the degree of consistency between observed and simulated patterns relative to various uncertainties and internal variability of the Earth system, and these conclusions are typically dichotomous (i.e., ``yes,'' there is a human influence, or ``no,'' there is not). Methods to attribute climatic trends without regression but assuming only additivity \citep{Ribes2017} are also available. The recent advent of large climate model ensembles now permits a straight forward analysis of how  distributions of climatic variables change when individual forcing agents are varied \citep{Angelil,Smith2022}. Such climate model-based attribution statements rely on the ``fitness for purpose'' of the phenomena in question. Note that climate models are far from perfect \citep{lee2024systematic} and such biases introduce uncertainty into an attribution statement.  Observations-only trend attribution statements can be also constructed, see \cite{Risser2024anthropogenic} and Section~\ref{sec:statCF}. 

Alternatively, event attribution is a more recent field of study that evaluates the human influence on individual weather and climate events, typically those that are extreme and have a large impact on human society \citep{national2016attribution}. Event attribution  does not  say climate change ``caused'' a specific event, but instead attempts to quantify changes in the likelihood (or probability) and severity (or magnitude) of certain classes of events. 
The World Weather Attribution project (\url{https://www.worldweatherattribution.org/}) uses both observations and available coupled ocean-atmosphere climate models to conduct rapid attribution statements in the aftermath of extreme weather events \citep{ascmo-6-177-2020}. 
Alternatively, the use of paired, atmosphere-only experiments with and without anthropogenic forcing of the surface boundary conditions \citep[see, e.g., the C20C+ D\&A project;][]{Stone2019experiment,Angelil} can investigate how the human influence varies from year to year. Some extreme events like hurricanes, severe storms and certain heatwaves are not well represented in long, coarse resolution climate model simulations. In this case, so-called storyline approaches \sti{utilizing imposed global warming methods} \newTxt{that study the effect of an imposed estimate of global warming on extreme events} (commonly referred to as pseudo-global warming) using high resolution weather forecasting methods \citep{Pall2017diagnosing, Patricola2018anthropogenic, BercosHickey2022anthropogenic} are a better alternative. However, these methods provide  an attribution statement only about the change in magnitude and not about the change in probability of the event \citep{Risser_Stone2017}. Event attribution studies using only observations have only recently been used \citep{Risser2017, Faranda2024} and are discussed in detail in Section \ref{sec:statCF}. As D\&A is ultimately a signal-to-noise problem, all attribution studies (trend or event) must carefully consider and account for observational and sampling uncertainty, model structural uncertainty, the effect of internal climate variability, and uncertainties associated with a given statistical methodology \citep{Paciorek2018quantifying, Delsole2019confidence, li2021uncertainty,Cummins2022}. 

\newTxt{While in principle D\&A methods can be applied to changes in spatial patterns, shifts, or variability, in most cases the focus is on changes over time. The focus on time trends and variability is natural since anthropogenic forcing has a clear time dependence. 
As such, there are a large number of approaches and methods relevant to D\&A in the area of econometrics and causal time-series analysis. For example, the economics literature uses econometrics methods that carefully decompose a time series into slowly evolving trend and cyclical terms in order to formally test for nonstationarity and the presence of a human influence on global temperatures \citep{Dergiades2016, Chang2020}. Alternatively, \cite{Pretis2020} draw an equivalence between energy-balance models of the climate and econometric models and show how the latter approach can be used to estimate climate responses and feedbacks. Another set of papers use econometrics techniques to generate counterfactuals from plausibly causal statistical models to argue for the causal effect of greenhouse gas emissions on socio-economic outcomes \citep{OrtizBobea2021, Diffenbaugh2019}. }


At its core, climate change D\&A fits within the broader field of causal inference, a collection of techniques used to identify cause and effect relationships. Causal inference methods first appeared in agriculture \citep{fisher1960design} before becoming widely used in the social sciences \citep{Sobel2000,Gangl2010}, biological sciences \citep{Kleinberg2011}, and epidemiology \citep{Rothman2005}. Within D\&A, traditional methods like optimal fingerprinting leverage the concept of \textit{Pearl causality} \citep{pearl2009causality, Hannart2015}, which uses  counterfactual theory to assess the influence of a specific ``intervention.'' Pearl-causal attribution leverages dynamical climate models, e.g., Coupled General Circulation Models (CGCMs), wherein one can apply interventions related to specific human activities (e.g., greenhouse gas forcing, sulfate aerosols, etc.) and compare with a corresponding simulation without the intervention. Such numerical experiments are freely available as part of the Detection and Attribution Model Intercomparison subproject \citep{gillett2016detection}, a part of the larger Coupled Model Intercomparison Project \citep{Eyring2016overview}. Model output is available via the Earth System Grid at \url{https://esgf-node.ornl.gov/}. The primary benefit of Pearl causality via counterfactuals is that a very clear causal statement can be made. However, GCM-based attribution conclusions are subject to model (or ``structural'') uncertainties and are only trustworthy when the models can accurately simulate the variable(s) of interest.
Alternatively, observations-only attribution that does not explicitly use counterfactuals relies on the concept of \textit{Granger causality} \citep{Granger1969}. Granger-based methods approach causality from a predictive standpoint and assess whether independent variables (or ``covariates'') provide better information about future values of a dependent variable relative to auto-regression. Granger-causal attribution statements are not as controlled as Pearl-causal statements since they do not involve a set of interventions, and are subject to confounding and mediation from  unaccounted-for covariates (see Section~\ref{sec:granger}). However, their primary reliance on observations means one can conduct attribution statements with very low computational cost and without worrying about whether the input data are ``representative'' of the real world (as is the case with climate models).  Whichever causal perspective is used, it is critical that climate change D\&A analyses clearly communicate their causal framing and how the resulting conclusions should be interpreted.

While there is now a clear definition and understanding for how Pearl causality applies to both trend \citep{Hasselmann1993optimal} and event attribution \citep{Hannart2015}, there is no analogous description of Granger-causal perspectives for climate change D\&A. Our goal in this review paper is therefore to provide a formal definition for Granger-based approaches to attribution and when such techniques are appropriate, as well as a clear comparison with Pearl-causal methods for D\&A. We provide a clear explanation of one commonly used Granger-based method, namely the \textit{statistical counterfactual} approach to trend and event attribution \citep{Risser2017,Risser2024anthropogenic}. The statistical counterfactual has now been used somewhat widely for attribution without a formal statement of assessment of its implied causal framing -- we intend to provide these foundational details.

The paper proceeds as follows. In Section~\ref{sec:granger}, we present a formal definition of Granger causality and its position within the broader field of causal inference. Section~\ref{sec:statCF} provides a formal definition for the statistical counterfactual approach to Granger-causal attribution \newTxt{with a clear demonstration of how it can be used to quantify and interpret a Granger-causal attribution statement.} \sti{as well as a clear comparison for how this method relates to more traditional attribution methods.} In Section~\ref{sec_sc_eg}, we provide an illustrative example of how the statistical counterfactual can be used to conduct Granger-causal D\&A for trends in global mean surface temperature. Section~\ref{sec:hybrid} describes a more recent hybrid approach to climate change D\&A that use both Granger and Pearl causality and provide a path forward for attribution conclusions that leverage the strengths of both approaches to D\&A. \newTxt{Section~\ref{sec:conclusions} concludes the paper.}

\section{Perspectives on Granger causality} \label{sec:granger}

\subsection{\newTxtR{Background and statistical notation}} \label{sec:granger:sub1}

Granger causality (GC) was proposed as a framework to detect causality in the late 1960s by the economist Clive \cite{Granger1969}. GC primarily relies on two assumptions: (1) the cause precedes the effect in time, and (2) the cause provides useful information for the prediction of the effect (outcome variable). It is because of these assumptions that GC is often referred to as ``predictive causality" since the notion of causality is closely tied to prediction. The first assumption entails that for a variable $X$ to be a cause of variable $Y$, it must precede it in time. This highlights one of the limitations of GC, namely its inability to detect contemporaneous (instantaneous) causality. Instantaneous causality refers to interactions between variables that occur within the sampling time step of observations or simulations. For example, the influence of precipitation on runoff in a small watershed may manifest within a few hours, constituting instantaneous causality if observations are solely available at a daily temporal resolution.  

The basic implementation of Granger causality relies on constructing two vector autoregressive (VAR) models, restricted and unrestricted. Throughout this section, we consider three types of variables: $X$, the causal variable(s) of interest; $Y$, the response or outcome variable of interest; and $Z$, variables that may \textit{mediate} or \textit{confound} the relationship between $X$ and $Y$.  We will revisit the importance of $Z$ later in this section. Without loss of generality, suppose we have measurements of each variable over time, e.g., $\{Y(t), X(t), Z(t): t = 1, \dots, T\}$. To test the hypothesis that variable $X$ is causing $Y$, first, a $p^{th}$-order unrestricted VAR model is developed: 
\begin{equation} \label{eq:GCunrestr}
    Y(t) = \mu + \sum_{i=1}^{p} \alpha_i Y(t-i) + \sum_{i=1}^{p} \beta_i X(t-i) + \sum_{i=1}^{p} \gamma_i Z(t-i) + \varepsilon(t).
\end{equation} 
In Equation~\ref{eq:GCunrestr}, $Y$ is expressed as a function of its own lagged values as well as lagged values of $X$ and $Z$; the formulation also includes an overall intercept $\mu$ and an error term $\varepsilon(t)$. 
Statistical techniques are used to estimate the coefficients $\{\alpha_i, \beta_i, \gamma_i: i = 1, \dots, p\}$ based on data.
Following the construction of the unrestricted model, a restricted model is developed:
\begin{equation} \label{eq:GCrestr}
    Y(t) = \mu^R + \sum_{i=1}^{p} \alpha^R_i Y(t-i) + \sum_{i=1}^{p} \gamma^R_i Z(t-i) + \varepsilon^R(t),
\end{equation} 
where again the restricted coefficients $\{\alpha^R_i, \gamma^R_i: i = 1, \dots, p\}$ are estimated using statistical methods.
The restricted model in Equation~\ref{eq:GCrestr} is similar to the unrestricted one with the exception of removing $X$ from the predictors. The null hypothesis that $X$ does not cause $Y$ is rejected if it is shown that the first model improves estimation compared to the second model according to an $F$-test. That is, \sti{causality is established if $X$ provides statistically-significant information for the prediction of $Y$} \newTxt{Granger causality is established if values of $X$ from previous time steps provide statistically significant information for the prediction of $Y$}.

\subsection{\newTxtR{Extensions and conditions for applying Granger causality}} \label{sec:granger:sub2}

\newTxtR{In addition to the aforementioned assumptions of time precedence and predictive causation,} the two VAR models in Equations~\ref{eq:GCunrestr} and \ref{eq:GCrestr} imply a third assumption of GC, namely that the underlying causal process is linear and stochastic \citep{Ombadi2020}. \sti{This assumption limits the applicability of GC in detecting causality in highly nonlinear systems.}
\sti{However,} \newTxt{For highly nonlinear systems,} methods based on information theory such as Transfer Entropy \citep[TE;][]{Schreiber2000} provide an extension of GC. \sti{to nonlinear systems.} TE is based on the concept of Shannon Entropy \citep{Shannon1948}, and causality is assessed by quantifying the reduction in the entropy of variable $Y$ if information about $X$ is known. For Gaussian processes, TE is equivalent to Granger causality \citep{Barnett2009}; however, TE is particularly advantageous in detecting causality in highly nonlinear systems. It should be noted that TE has not often been used in climate change attribution science, and it holds significant potential, with the caveat that it has been shown to require a significantly larger sample size compared to GC \citep{Ombadi2020}. 

\sti{In addition to the abovementioned three assumptions of GC (time precedence, predictive information of the cause, and linearity), an important} \newTxt{The final} assumption of GC is separability \newTxt{of causal factors}. That is, the information contained in the causal variable $X$ is unique and can be completely removed from the system by eliminating $X$ in the restricted model of Equation~\ref{eq:GCrestr}. While this assumption is satisfied in stochastic, linear systems, it is an invalid assumption in deterministic systems with weak to moderate coupling \citep{Sugihara2012}. In such systems, information about the causative factor $X$ are contained in past values of the effect $Y$; therefore, the restricted model can not be used to eliminate the effect of $X$ and assess predictive causality. The method of Convergent Cross Mapping \citep[CCM;][]{Sugihara2012} has been proposed as an alternative paradigm of causality that can be used in such settings of deterministic, dynamical systems. For instance, CCM has been used to disentangle the causal feedback between temperature and CO$_{2}$ concentrations during Earth glacial cycles \citep{vanNes2015}. 

\newTxt{We have so far introduced four major assumptions for Granger causality: time precedence, predictive information of the cause, linearity, and separability.}
Out of these four assumptions, two are absolute requirements: predictive information of the cause and separability. The other two assumptions are corollaries of the standard implementation of GC. For example, nonlinear terms (e.g., $X^2$ or $\exp\{X\}$), nonlinear versions of the two models (e.g., multiplicative instead of additive), or even interaction terms (e.g., $X\times Z$) could be considered, although the latter is challenging within the framework of GC. \newTxtR{In certain specific circumstances, such as when dealing with aggregated quantities over longer time scales,} \st{the time precedence assumption can be relaxed by including} ``instantaneous'' causality \newTxtR{can be permitted by adding non-lagged quantities to} the right-hand side of Equations~\ref{eq:GCunrestr} and \ref{eq:GCrestr}. This may only be appropriate when the time scales of interest correspond to aggregated quantities, e.g., seasonal or annual means (see Section~\ref{sec_sc_eg}). \newTxtR{In this case, time precedence is still present but is represented implicitly instead of explicitly.}

\newTxt{Finally, in order for Granger-causality methods to be applied to a given data set, one must ensure that all input variables ($X$, $Y$, and $Z$) are stationary \citep{Granger1969}. In other words, each variable should have a constant mean, constant variance, and no cyclical behavior. A basic result from time series econometrics is that any regression of non-stationary variables (e.g., nonstationary $Y$ regressed on nonstationary $X$) will most likely yield a statistically significant relationship, even though no causal meaning can be attributed to this relationship. 
\newTxtR{For example, when assessing the underlying causal factors for changes to global mean surface temperature (which is primarily increasing over the last century; see Section~\ref{sec_sc_eg}), one could likely obtain a statistically significant regression coefficient for any monotonically increasing variable $X$, even those that have no relationship with global temperatures, e.g., ice cream sales or the number of pet dogs.}
Formal statistical tests can be used to determine whether each input variable is stationary, e.g., the augmented Dickey-Fuller test \citep{Elliott1996} or the Kwiatkowski–Phillips–Schmidt–Shin test \citep{Kwiatkowski1992}. If the variables of interest are deemed to be nonstationary, a common approach is to instead analyze the first- or second-order differences. See Section~\ref{sec_sc_eg} for more details.}

\subsection{\newTxtR{Mediation and confounding}} \label{sec:granger:sub3}

\begin{figure}[!t]
    \centering
    \includegraphics[trim={85 100 105 10mm}, clip, width=\textwidth]{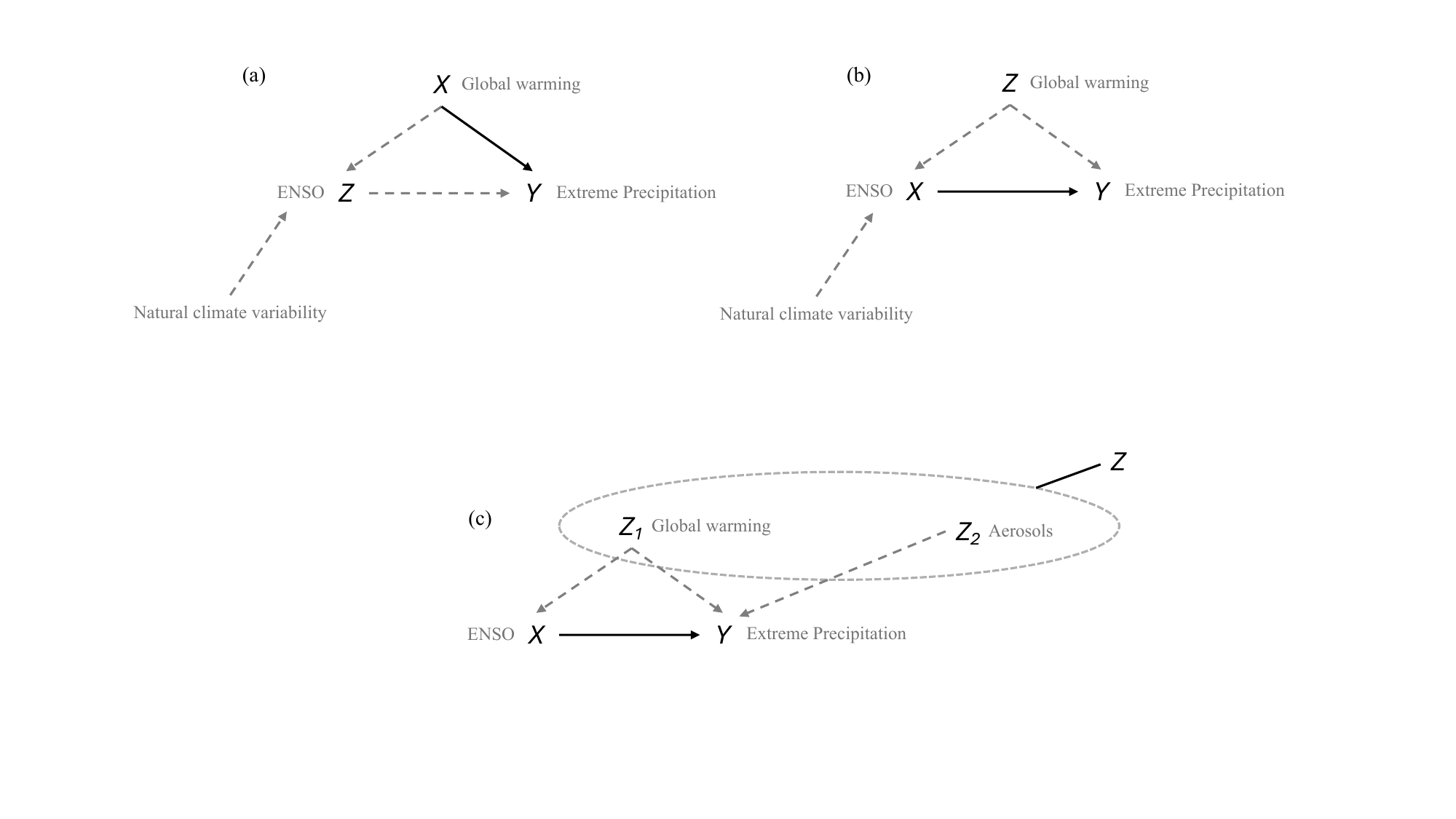}
    \caption{Causal graphs illustrating the relationship between the causal variable $X$, response or outcome variable $Y$, and covariate(s) $Z$ used in Equations ~\ref{eq:GCunrestr} and \ref{eq:GCrestr}.
    \newTxt{In all panels, solid lines indicate the causal relationship being tested, while the dashed lines represent additional causal relationships within the system.}
    Panel (a) shows the hypothesis that global warming ($X$) causes an increase in the intensity of extreme precipitation ($Y$), which must take into consideration removing the effect of natural variability such as ENSO ($Z$). In this example, $Z$ acts both as a mediator through which global warming indirectly affects extreme precipitation and an independent causal factor representing the impact of natural variability. In panel (b), the hypothesis is concerned with testing the impact of ENSO ($X$) on extreme precipitation ($Y$). Here, it is important to consider removing the effect of global warming ($Z$), which is a confounding variable, causing both $X$ and $Y$. Panel (c) shows a causal graph similar to that of (b), including an additional independent causal factor, namely aerosols.}
    \label{fig_confounders}
\end{figure}

As previously stated, the $Z$ variable(s) are statistical covariates that may \textit{mediate} or \textit{confound} the relationship between $X$ and $Y$. \textit{Mediation} refers to a variable lying between $X$ and $Y$ on the causal pathway, i.e., $X \rightarrow Z \rightarrow Y$, where ``$\rightarrow$'' indicates causation. \textit{Confounding} refers to causation of both $X$ and $Y$, i.e., $X \leftarrow Z \rightarrow Y$. Of course, $Z$ may also be an \textit{independent} causal factor, wherein $Z$ and $X$ are unrelated while $X \rightarrow Y$ and $Z \rightarrow Y$. Figure~\ref{fig_confounders} shows example causal graphs associated with each of these cases along with corresponding attribution hypotheses. 
\st{Figure 2a shows an example of an attribution hypothesis in which $Z$ is both a mediator and an independent causal factor. The hypothesis involves testing whether global warming is directly increasing the intensity of extreme precipitation. In order to properly test this hypothesis using GC, one must include ENSO to account for the impact of global warming on extreme precipitation that is mediated through ENSO. Furthermore, including ENSO in $Z$ is important to remove the effects of natural variability on extreme precipitation.}
\newTxtR{Figure \ref{fig_confounders}a illustrates an attribution hypothesis regarding whether global warming is increasing the intensity of extreme precipitation. Here, $Z$ (ENSO) is both a mediator and an independent causal factor. If the hypothesis seeks to evaluate the \textbf{direct} effect of global warming on intensifying extreme precipitation--without considering any indirect effects mediated by ENSO--then ENSO should be treated as a $Z$ variable in the GC model. Similarly, if there is reason to believe that natural climate variability, correlated with global warming, contributes to the variation, ENSO should be included in $Z$. In all other cases, controlling for ENSO should be avoided, as it leads to a biased estimate of the total causal effect ($X \rightarrow Y$).}
On the other hand, Figure~\ref{fig_confounders}b shows an example of an attribution statement in which $Z$ is a confounder. Here, the aim is to test whether ENSO increases the intensity of extreme precipitation, which requires removing the impact of global warming on both $X$ and $Y$. 
\st{Finally, Figure 2c shows an example in which $Z$ includes two variables, namely global warming and aerosols. The former is a confounder while the latter is an independent causal variable that must be accounted for in GC. Because the causal graphs for a given hypothesis are not known a priori, selecting the variables to include in the set $Z$ can be extremely challenging and often requires substantial domain knowledge of the research problem at hand.  Furthermore, appropriately identifying $Z$ is critical for GC inferences to be reliable: if any confounders are omitted from the model, the analysis could produce misleading results.}
\newTxtR{Finally, Figure~\ref{fig_confounders}c illustrates an example in which $Z$ includes two variables: global warming and aerosols. The former is a confounder, while the latter is an independent causal variable that must be accounted for in GC. Since the causal graphs for a given hypothesis are not known \textit{a priori}, selecting the appropriate variables for $Z$ can be extremely challenging and often requires substantial domain knowledge of the research problem. Additionally, properly identifying $Z$ is crucial for ensuring the reliability of GC inferences. Specifically, $Z$ must include variables that co-vary with $X$; if such variables are omitted from the model, the analysis could produce misleading results.}
\newTxt{We discuss this further in Section~\ref{sec:hybrid}.}

\subsection{\newTxtR{Causal framing and comparison with Pearl causality}} \label{sec:granger:sub4}

Ultimately, the main distinction between GC and Pearl causality lies in their foundational paradigms. Pearl causality operates within a counterfactual framework, asking: \textit{what would the outcome be if the cause had not occurred?} In climate change attribution, counterfactual scenarios can be established using climate models; for instance, simulations of mean global surface temperature could be generated for different scenarios in which one can exclude or include a given driving mechanism (e.g., aerosols) to establish causality \citep{hegerl1997multi}. Counterfactual reasoning is only feasible if a model of the underlying system exists in the form of a mathematical equation or physical model. On the other hand, Granger causality does not establish causality in the strict philosophical sense but rather identifies predictive relationships between variables within a statistical framework. In summary, while both Granger and Pearl causality are used to assess causality, Pearl causality provides a more principled and rigorous framework for causal inference, particularly in complex systems such as the climate system where causal relationships may be influenced by multiple factors. \newTxtR{In either framework, the models used should always be supported by our understanding of the physical mechanisms driving changes to the Earth system.}

\section{Statistical counterfactuals} \label{sec:statCF}


\subsection{\newTxtR{Motivation and formal definition}} \label{sec:statCF:sub1}

The main emphasis of \st{both Pearl and} Granger-based approaches to climate change attribution is a hypothesis test \st{(or tests)} to determine causality. Granger causality uses an $F$-test to compare the unrestricted and restricted autoregressive models, 
\st{while optimal fingerprinting (the prominent Pearl-causal method for trend attribution) conducts a pair of tests to first \textit{detect} a meaningful trend and subsequently \textit{attribute} trends to specific forcing agents. For optimal fingerprinting, the hypothesis tests involve the scaling factor for a given forcing agent, a unitless quantity that can be difficult to interpret when it deviates from unity. In both cases,}
\newTxtR{which outputs a} \st{the output of a hypothesis test is a} $P$-value that summarizes the strength of evidence against the null hypothesis of interest; \newTxtR{the $P$-value} is then converted to a dichotomous yes/no conclusion regarding causality. \st{Ultimately, neither of these approaches yield straightforward results  } \newTxtR{This procedure and the resulting conclusion make it difficult} to quantify or interpret the causal relationship of interest, and it is not directly obvious how to compare or rank the relative importance of multiple causal variables of interest.
The \textit{statistical counterfactual} approach to Granger-causal trend and event attribution, formally defined by \cite{Risser2017}, is one approach that is useful for quantifying the magnitude of a causal relationship while interpreting and intercomparing the effects of multiple causal factors \citep[see also][for a different flavor of this methodology that was developed separately]{Vautard2019, ascmo-6-177-2020}. 

For climate change attribution (either trend or event), a statistical counterfactual analysis with Granger causality is based on the unrestricted autoregressive model in Equation~\ref{eq:GCunrestr}, where the response or dependent variable $Y$ is derived from observations. 
Here, ``observations'' could refer to measurements from weather gauges \citep[e.g.][]{Menne2012}, gridded data products \citep[e.g.,][]{Livneh2014}, reanalysis products \citep[e.g.,][]{Hersbach2020era5}, or even satellite measurements. 
\newTxt{Generally speaking, in a Granger analysis the $X$ and $Z$ variables are also derived from observations (although this need not strictly be the case; see Section~\ref{sec_sc_eg}, \newTxtR{where the anthropogenic aerosol forcing time series is derived from models}).} \st{, which is in distinct contrast to optimal fingerprinting where the patterns (the $X$) are model-derived.}
As in Section~\ref{sec:granger}, suppose we have measurements of an observed climate variable of interest indexed over time, i.e., $\{Y(t): t=1, \dots, T\}$, which could represent measurements from a single geospatial location or the spatial average of a set of weather stations or grid boxes. The time units typically refer to an annual or seasonal quantity, e.g., the average or maximum of daily or monthly data. \newTxt{For annually- or seasonally-aggregated quantities, we often include the ``instantaneous'' aggregated values of the causal variable(s) $X$ and confounder(s) $Z$ in the autoregressive models since these encode the most recent history of the process of interest.} \newTxtR{(Again, in this case the time dependence is implicit for the time-aggregated instantaneous variables.)} The goal of the statistical counterfactual approach is to quantify changes in the statistics or \textit{distribution} of $Y(t)$ over time as specified by the unrestricted model.
\newTxtR{Here and throughout the paper, the term ``counterfactual'' is used to refer to a climate scenario that has never occurred before and therefore can only be reconstructed statistically or using physical models. The approach described here approximates counterfactuals statistically:}
conditioned on the inferred coefficients \newTxtR{from the unrestricted model} and their uncertainty, we can estimate a counterfactual distribution for the response $Y(\cdot)$ for an arbitrary combination of inputs.

First, note that the unrestricted autoregressive model in Equation~\ref{eq:GCunrestr} supposes that the \textit{expected value} or mean behavior of $Y(t)$, denoted $m(t) = E[Y(t)]$, can be written as
\begin{equation} \label{eq:statCF}
    m(t) = \underbrace{\mu}_{\hbox{\small Pre-ind.}} + \underbrace{m_Y(t)}_{\hbox{\small Recent past}} + \underbrace{m_X(t)}_{\hbox{\small Ext. forced}} + \underbrace{m_Z(t)}_{\hbox{\small Internal var.}}.
\end{equation}
(Here we have assumed that we only have a single $X$ and a single $Z$; this can of course be generalized; see Section~\ref{sec_sc_eg}.) In  Equation~\ref{eq:statCF}, $\mu$ represents the time-invariant, pre-industrial (or background) climatology, unperturbed by human influence;
$m_Y(t) = \sum_{i=1}^p\alpha_iY(t-i)$ describes the influence of the recent past in the response; $m_X(t) = \sum_{i=1}^p\beta_iX(t-i)$ and $m_Z(t) = \sum_{i=1}^p\gamma_iZ(t-i)$ characterize the influence of the causal and mediating variables, respectively. 
For climate change attribution, it is most common for the $X$ variables to represent time-evolving external forcing agents that correspond to anthropogenic activities, e.g., greenhouse gas emissions, anthropogenic aerosols, or land-use/land-cover change. Similarly, in most cases the $Z$ variables represent the influence of natural or internal climate variability, e.g., large-scale modes of variability such as the El Ni\~no/Southern Oscillation or the Pacific-North American teleconnection pattern. The $X$ and $Z$ variables are sometimes referred to as ``covariates'' (terminology borrowed from the statistical literature); these are typically considered fixed and known, although one could assume they involve uncertainty -- e.g., $X(t) = X^*(t) + e_X(t)$, where $X^*(t)$ is the unknown ``true'' value contaminated by noise $e_X(t)$ \citep[see, e.g.,][]{Lau2023extreme}. 

Equation~\ref{eq:statCF} becomes a statistical model by assuming the data $Y(t)$ will deviate unpredictably from the average behavior $m(t)$ via the error term $\varepsilon(t)$. The error $\varepsilon(t)$ accounts for the fact that some of the year-to-year internal variability in a given climate variable cannot be described by forced changes and large-scale natural climate drivers, typically arising from short-term weather variability associated with a chaotic atmosphere and ocean dynamics, among other things. One furthermore assumes that the deviations $\varepsilon(t) = Y(t) - m(t)$ are \textit{random variables} that follow a given statistical distribution.  \sti{and are statistically independent.} \newTxt{In most cases, the deviations are assumed to be statistically independent, but one can also account for dependent errors.} The specific distribution that should be used depends on the nature of the underlying data $Y(t)$ and how one expects the deviations to behave: for example, if $Y(t)$ represents a seasonal or annual mean, $\varepsilon(t)$ might follow a Gaussian distribution; alternatively, if $Y(t)$ represents a seasonal or annual maximum, then $\varepsilon(t)$ might follow a Generalized Extreme Value distribution. The error distribution will also depend on statistical parameters that are considered unknown and must be inferred from the data; these parameters can be time-varying and, like the average behavior $m(t)$, can depend on covariates \citep[see, e.g.,][]{Risser2017}. Ultimately, the uncertainty in $\varepsilon(t)$ is an estimate of the internal variability innate to $Y$, above and beyond that which is accounted for by the $X$ and $Z$ variables. 

\subsection{\newTxtR{Defining counterfactuals statistically}} \label{sec:statCF:sub2}

The statistical counterfactual approach uses the fact that, \newTxt{within the assumptions of the unrestricted model}, the covariates $X$ and $Z$ describe all \sti{secular} \newTxt{long-term} trends and year-to-year natural variability in the underlying distribution of the data $Y(t)$ (above and beyond the influence of the recent history of the data). 
Because the statistical distribution of the data is time-varying (in terms of the covariates), we can generate time-varying estimates of arbitrary statistical summaries from the fitted statistical model.
This is accomplished by plugging in a particular combination of covariates to yield estimates of, e.g., the expected value \newTxtR{(see Section~\ref{subsec:interp})}, an upper tail quantile of the response variable $Y$, or the probability that $Y$ will exceed a given threshold \newTxtR{(see Section~\ref{subsec:interp:ea})}. 
In addition to an arbitrary combination of covariates, the coefficients
$\{\alpha_i, \beta_i, \gamma_i: i = 1, \dots, p\}$ \newTxtR{and the parameters associated with the error term $\varepsilon(\cdot)$}
from Equation~\ref{eq:GCunrestr} determine the time-varying distribution. Since we are again most often interested in the effect of the causal variable $X$, the main emphasis is on its coefficients $\{ \beta_i\}$,
which we refer to as ``statistical betas'', a nod to the beta terms in standard linear regression attribution methods \citep{Allen2003estimating}. The statistical betas are regression coefficients that quantify the relationship between $X$ and $Y$; their units are the change in $Y$ per unit increase in $X$. Given a finite amount of data, the statistical betas involve uncertainty, which must be appropriately quantified and accounted for when determining the statistical significance of any attribution conclusions. 

When the combination of covariate values used correspond to conditions that occurred in the real world (e.g., from a specific year), the resulting estimates summarize our best estimates of the distribution of the data from that year. The ``counterfactual'' aspect is introduced when we plug in a set of covariate values that did not actually occur in reality: for example, externally-forced covariates from a pre-industrial climate together with present-day modes of natural climate variability. The statistical counterfactual can thus be used to isolate the effect of an individual covariate (externally-forced or natural driver) or even a set of covariates while holding all other conditions constant. Ultimately, attribution conclusions are obtained by conducting a hypothesis test for statistical significance of a metric of change relevant to the problem at hand: this could be a mean change in trend attribution (see Section~\ref{subsec:interp}), or a change in a magnitude (return level) or probability (i.e., the risk ratio; see Section~\ref{subsec:interp:ea}) for event attribution. 

In order to demonstrate the importance of the statistical betas, let us briefly consider a simplified version of the unrestricted model that has no autoregressive terms and leverages the instantaneous values of $X$ and $Z$. That is, Equation~\ref{eq:statCF} becomes $m(t) = \mu + \beta X(t) + \gamma Z(t)$
(with no autoregression, $m_Y(t)$ is dropped from the equation). Furthermore, suppose that we are interested in quantifying the \st{maximum} effect of the causal factor $X$ \newTxtR{in trend attribution}, denoted $\Delta_X$, while holding the ``background'' conditions specified by $Z$ constant.
\st{The null hypothesis is that of no change, i.e., $H_0: \Delta_X = 0$, compared with the alternative that the change is nonzero, i.e., $H_1: \Delta_X \neq 0$.}
To assess these hypotheses, we can generate two counterfactual scenarios:
\begin{enumerate}
    \item $X(t_{+})$, where $t_{+} = \arg\max X(t)$, with the confounding variable set to its climatological average $\overline{Z}=\frac{1}{T}\sum_t Z(t)$, compared with
    \item $X(t_{-})$, where $t_{-} = \arg\min X(t)$, and the confounding variable again set to $\overline{Z}=\frac{1}{T}\sum_t Z(t)$.    
\end{enumerate}
The mean (expected value) of the data for each scenario is
\[
m_1 = \mu + \beta X(t_{+}) + \gamma \overline{Z}, \hskip3ex \text{and} \hskip3ex m_2 = \mu + \beta X(t_{-}) + \gamma \overline{Z},
\]
which we can use to calculate the maximum effect of $X$ as
\[
\Delta_X = m_1 - m_2 = (\mu + \beta X(t_{+}) + \gamma \overline{Z}) - (\mu + \beta X(t_{-}) + \gamma \overline{Z}) = \beta \big[X(t_{+}) - X(t_{-})\big]
\]
In other words, the change $\Delta_X$ is simply a rescaling of $\beta$, meaning that testing the null hypothesis $H_0: \Delta_X = 0$ is equivalent to testing the null hypothesis $H_0: \beta = 0$, i.e., our attribution conclusions completely depend on the statistical betas associated with this approach. Importantly, in estimating $\Delta_X$, the statistical counterfactual effectively isolates the effect of the causal variable $X$ by removing the influence of the mediating variable(s) $Z$. However, since the $\beta$ terms were estimated systematically with the $Z$ variables included in the model, any mediating effects of $Z$ are still accounted for.

In the above example, note that testing $H_0: \Delta_X = 0$ is equivalent to the $F$-test comparing the unrestricted and restricted models: $\Delta_X=0$ if and only if all $\beta_i=0$, in which case the unrestricted model becomes the restricted model. 
One common case where this equivalence can not be made is in the case of extreme event attribution, where the change of interest is often the risk ratio (i.e., a ratio of probabilities for the two scenarios of interest). 
Similar to the example above, the statistical counterfactual could similarly generate estimates of return probabilities as a function of the coefficients $\{\alpha_i, \beta_i, \gamma_i: i = 1, \dots, p\}$ and \newTxtR{parameters associated with the error term $\varepsilon(\cdot)$} \citep[see, e.g.,][]{Coles2001}. However, return probabilities are a nonlinear function of the coefficients and hence the risk ratio cannot be written as a function of only the statistical betas, one must therefore directly test the null hypothesis $H_0: RR = 1$ \citep{Paciorek2018quantifying}.

\subsection{\newTxtR{Examples from the trend and event attribution literature}} \label{sec:statCF:sub3}

\sti{Since its formal introduction in Risser et al. (2017) for use in event attribution,} The statistical counterfactual has been used for a variety of attribution studies: \newTxt{for example, \cite{Risser2017} assessed the human influence on the heavy rainfall experienced during Hurricane Harvey;} \cite{van2019human} conducted event attribution for the European heatwave in June, 2019; \cite{Keellings2019} assessed the human influence on rainfall in Puerto Rico associated with Hurricane Maria; \cite{Risser2021quantifying} isolated the effect of natural modes of climate variability on mean and extreme precipitation in the United States; \cite{BercosHickey2022anthropogenic} and \cite{zhang2024explaining} conducted event attribution for the 2021 Pacific Northwest heatwave; \cite{Risser2024anthropogenic} conducted trend attribution for seasonal  mean and extreme precipitation in the United States; \cite{Risser2024impossible} assessed the human influence on statistically ``impossible'' temperatures. 
We reiterate that the statistical counterfactual is built upon the framework of Granger causality, and as
such, its causal framing is the same as the Granger perspective described at the end of Section~\ref{sec:granger}: the focus is on \textit{predictive} causality that may be obscured by hidden covariates. While this is a weaker form of causality relative to intervention-based counterfactuals used with Pearl causality, we reiterate that Granger-causal attribution is nonetheless important, particularly when used in tandem with Pearl statements. Our overall understanding of how human activity influences the climate system and our confidence in the evidence for attribution is increased when both methods yield similar conclusions.

\section{Example: changes to global mean surface temperature} \label{sec_sc_eg}

\sti{In order to make these concepts clear,} \newTxt{In order to illustrate the use of the statistical-counterfactual method,} we now provide a brief example. \st{Suppose we are interested in trend attribution of global mean surface temperature (GMST) due to human-induced changes in atmospheric radiative forcing over the last century.}
\newTxtR{Suppose we are interested in determining the impact of human-induced changes in atmospheric radiative forcing on global mean surface temperature (GMST) over the last century. Specifically, we want to conduct both trend attribution of long-term changes to GMST as well as determine the extent to which anthropogenic forcing influenced the probability of experiencing a GMST at least as extreme as 2015 (which saw a GMST of $0.9^\circ$C above the 1950-1980 average). Both of these questions can be answered using Granger causality and the statsitical counterfactual approach described in Section~\ref{sec:statCF}.}

\begin{figure}[!t]
    \centering
    \includegraphics[trim={0 0 0 0mm}, clip, width=\textwidth]{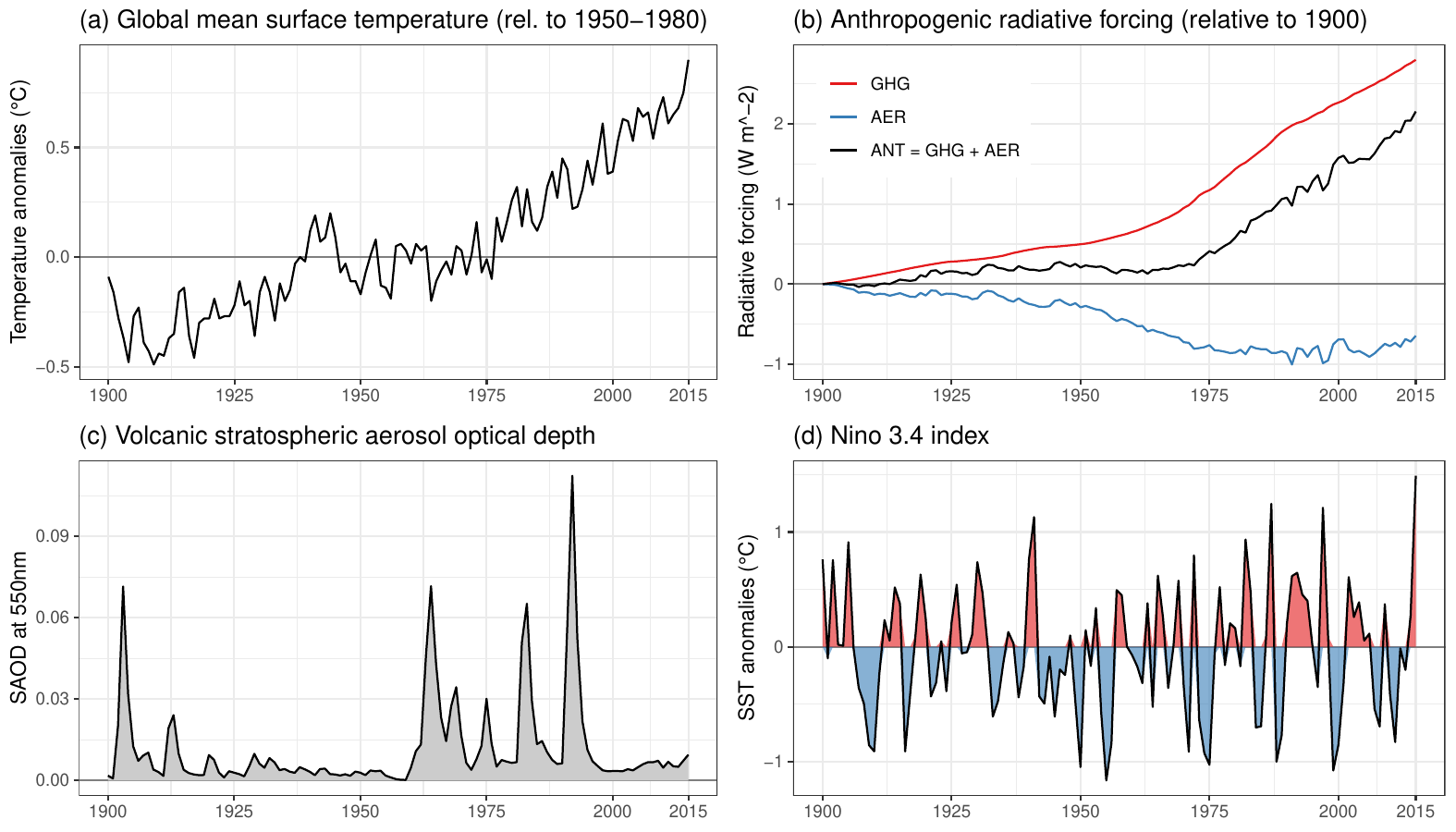}
    \caption{Input data for the statistical counterfactual analysis described in Section~\ref{sec_sc_eg}. Panel (a) shows global mean surface temperature anomalies ($^\circ$C) relative to 1950-1980 \citep[from][]{Lenssen2019improvements}; panel (b) shows radiative forcing (W m$^{-2}$) from greenhouse gas forcing, anthropogenic aerosols, and the combined anthropogenic forcing; panel (c) shows stratospheric aerosol optical depth from volcanic activity; panel (d) shows Ni\~no3.4 index anomalies.}
    \label{fig_example}
\end{figure}

To answer \st{this question} \newTxtR{these questions}, let $Y(t), t = 1900, \dots, 2015$ be annual averages of monthly GMST measurements relative to 1950-1980 \citep[obtained from][]{Lenssen2019improvements}. For simplicity, suppose we 
are interested in the causal effects of a single variable that represents the sum-total atmospheric radiative forcing from greenhouse gas emissions and anthropogenic aerosols (denoted ``ANT''), i.e., $X(t) = \text{GHG}(t) + \text{AER}(t)$, where $\text{GHG}(t)$ is the radiative forcing due to greenhouse gas emissions in year $t$ \citep[as used in, e.g.,][]{Risser2024anthropogenic} and $\text{AER}(t)$ is the global radiative forcing due to anthropogenic aerosol forcing in year $t$ \citep[taken from][]{Smith2021,Smith2021data}. Here we use the sum-total anthropogenic forcing as the covariate because otherwise GHG and aerosol forcing are not separable (i.e., they are strongly anti-correlated or ``collinear''). We also account for two mediation variables which are known to drive chages to global temperatures: the effect of volcanic stratospheric aerosol forcing, denoted $Z_\text{Volc}(t) = \text{vSAOD}(t)$, where $\text{vSAOD}(t)$ is the stratospheric aerosol optical depth due to volcanic activity in year $t$ \citep[obtained from][]{sato1993stratospheric, schmidt2018volcanic}; and the Ni\~no 3.4 index (obtained from \url{https://psl.noaa.gov/data/timeseries/month/DS/Nino34/}), denoted $Z_\text{ENSO}(t)$, which describes the effect of the El Ni\~no/Southern Oscillation (ENSO) on global temperatures. Both of these confounding variables are natural, i.e., non-anthropogenic, although volcanic activity is often deemed an ``external'' forcing to the climate system while ENSO is a mode of variability that is ``internal'' to the climate system (and often considered an aspect of internal climate variability). Like the global temperature variable, $X(t)$ and the $\{Z_{(\cdot)}(t)\}$ are annually-averaged. These variables are plotted in Figure~\ref{fig_example}.

\newTxtR{In order to conduct both trend and event attribution using the statistical counterfactual, one must first assess Granger causality using the formal techniques described in Section~\ref{sec:granger}. Assuming Granger causality can be established, the statistical counterfactual can proceed to provide both trend and event attribution conclusions. }
All of the data analysis summarized in the next \st{two} \newTxtR{three} sections is conducted using standard statistical software for linear regression with Gaussian errors (code provided in the Supporting Information).

\subsection{Assessing Granger causality} \label{sec_example_sub1}

The first step associated with conducting a Granger-causal statement for these data is to assess whether the different input variables are stationary. One widely used hypothesis test for determining stationarity is the augmented Dickey-Fuller (ADF) test \citep{Elliott1996}, which tests the null hypothesis that a unit root is present in a time series, i.e., that the time series is nonstationary. In our case, the alternative is that the data are stationary; small $P$-values indicate that the null hypothesis can be rejected and the data can safely be assumed to be stationary. Supplemental Table S1 shows the results of applying the ADF test to each variable. Unsurprisingly, the ADF test concludes that the global mean surface temperature and sum-total anthropogenic radiative forcing variables are not stationary. However, if we apply the test to the first-order differences, denoted
\[
Y^*(t) = Y(t)-Y(t-1), \hskip1ex X^*(t) = X(t)-X(t-1), \hskip1ex Z_{(\cdot)}^*(t) = Z_{(\cdot)}(t)-Z_{(\cdot)}(t-1), 
\]
we instead find that all variables are decidedly stationary. Hence, we proceed with the analysis using first-order differences of all variables.

We then consider two statistical models: first, the unrestricted $p^\text{th}$-order autoregressive model
\begin{equation} \label{eq:app:unr}
Y^*(t) = \mu + \sum_{i=1}^{p} \alpha_i Y^*(t-i) + \sum_{i=0}^{p-1} \beta_i X^*(t-i) + \sum_{i=0}^{p-1} \gamma_{\text{Volc},i} Z^*_{\text{Volc}}(t-i)+ \sum_{i=0}^{p-1} \gamma_{\text{ENSO},i} Z^*_{\text{ENSO}}(t-i) + \varepsilon(t);
\end{equation}
second, the restricted $p^\text{th}$-order autoregressive model
\begin{equation} \label{eq:app:res}
Y^*(t) = \mu + \sum_{i=1}^{p} \alpha^R_i Y^*(t-i) + \sum_{i=0}^{p-1} \gamma^R_{\text{Volc},i} Z^*_{\text{Volc}}(t-i)+ \sum_{i=0}^{p-1} \gamma^R_{\text{ENSO},i} Z^*_{\text{ENSO}}(t-i) + \varepsilon(t).
\end{equation}
Note that here we have included the ``instantaneous'' anthropogenic radiative forcing, volcanic aerosol, and ENSO state since we are looking at annually-aggregated quantities. 

Before conducting an $F$-test comparing the unrestricted and restricted models, we need to select the best value of $p$, i.e. the optimal number of lagged covariate values to include in the regression. A simple way to make this choice is to fit the unrestricted model many times across different values of $p$ and assess model fit using, e.g., Akiake's Information Criteria \citep[AIC;][]{akaike1998information}. These results are shown in Supplemental Figure S1(a), where we look at $p \in \{ 1, \dots, 15 \}$; $p=1$ corresponds to no autoregression. From this plot, including five lagged values (current value plus four previous) provides the best fit to the data.

Using the optimal number of time lags, we then fit both the unrestricted and restricted models to our data using standard statistical software. Supplemental Figure S1(b) shows the autoregressive coefficients for each variable under the unrestricted model. The $F$-statistic for testing Granger causality of anthropogenic radiative forcing on global mean surface temperature is
\[
F = \frac{(SSE_\text{rest.} - SSE_\text{unrest.})/(df_\text{rest.} - df_\text{unrest.})}{SSE_\text{unrest.}/df_\text{unrest.}},
\]
where $SSE_{(\cdot)}$ is the error sum of squares and $df_{(\cdot)}$ is the degrees of freedom for each model. Statistical theory tells us that the $F$-statistic follows a $F$ distribution with $(df_\text{rest.} - df_\text{unrest.})$ and $df_\text{unrest.}$ degrees of freedom. For this example, $F=2.08$, $df_\text{rest.} = 99$, and $df_\text{unrest.}=94$; the corresponding $P$-value is $0.07$. {Thus, using a significance threshold of $\alpha = 0.1$, we conclude that anthropogenic radiative forcing Granger-causes changes in global mean surface temperature.} \newTxtR{We choose a significance threshold of $\alpha = 0.1$ to correspond to the ``very likely'' confidence language used by the 
Intergovernmental Panel on Climate Change \citep{mastrandrea2010guidance}.}

Before proceeding, note that we can also assess Granger causality for changes in global mean surface temperature for volcanic aerosols and ENSO: simply swap the anthropogenic radiative forcing to be one of the confounding $Z$ variables and label volcanic aerosols or ENSO the $X$ variable of interest (one at a time). Using the same unrestricted model fit, we can then fit restricted models by first leaving out just volcanic aerosols and then leaving out just ENSO. This allows us to conclude that both volcanic aerosols ($F=7.72$; $P<0.001$) and the El Ni\~no/Southern Oscillation ($F=21.86$; $P<0.001$) Granger-cause changes in global mean surface temperature. Here, the evidence is stronger, such that the conclusion is significant for any reasonable significance level.

Interestingly, the $P$-values for ENSO and volcanic aerosols are much smaller than for anthropogenic forcing. Note that this does not imply that ENSO and volcanic aerosols are the primary drivers of global warming -- merely that their first-order differences are more strongly correlated with first-order differences in global temperature. \st{Indeed, when conducting the same analysis on the original measurements (not first-order differences), we find that anthropogenic radiative forcing has the smallest $P$-value (see Table 1). Furthermore, in both analyses (first-order differences or original measurements), we find that anthropogenic radiative forcing has a much larger effect on temperature measurements (described in Section 4.2; again see Table1). Note that the results using the original measurements are only for comparison: only the results using first-order differences are used to draw conclusions about the causal effects of each variable on global mean surface temperature.}
\newTxtR{One major reason for the lower confidence in attribution to anthropogenic forcing (relative to ENSO and volcanic aerosols) is its identifying variation after conversion to first-order differences. Since anthropogenic forcing tends to vary quite smoothly from year to year, there is little variation in the first-order differences from which to identify effects on temperature fluctuations. Indeed, as shown in Supplemental Figure S2, the first-order differences in the ANT time series in Figure~\ref{fig_example}(b) are less than $\pm10\%$ of the range of the original ANT values (compared with ENSO and volcanic aerosols, which have relative variation of more than $\pm60\%$ and $\pm40\%$, respectively). This highlights the challenges of identifying causation for covariates that vary smoothly (and nearly monotonically) over time, and is a limitation of the statistical counterfactual methodology.}

\subsection{Quantifying and interpreting causality: trend attribution} \label{subsec:interp}

\newTxtR{Now that we have established Granger causality for all three covariates of interest, we now proceed to use the statistical counterfactual approach to conduct trend attribution for each variable.}
As described in Section~\ref{sec:statCF}, we can use the fitted coefficients from the unrestricted model to summarize the effect of each causal variable on global mean surface temperature and rank the relative importance of each causal factor in driving changes in temperature. First, for clarity, we re-write the unrestricted model as
\[
Y^*(t) = \mu + m^*_{Y}(t) + m^*_{X}(t) + m^*_{Z_\text{volc}}(t) + m^*_{Z_\text{ENSO}}(t) + \varepsilon(t),
\]
where
\[
m^*_{Y}(t) = \sum_{i=1}^{p} \alpha_i Y^*(t-i), \hskip2ex m^*_{X}(t) = \sum_{i=0}^{p-1} \beta_i X^*(t-i), \hskip2ex \text{and} \hskip1ex m^*_{Z_{(\cdot)}}(t) = \sum_{i=0}^{p-1} \gamma_{(\cdot),i} Z^*_{(\cdot)}(t-i).
\]
Conditioned on estimates of the coefficients $\{\mu, \alpha_i, \beta_i, \gamma_{\text{Volc},i}, \gamma_{\text{ENSO},i}\}$ and their uncertainties, we could then construct estimates of the $m^*_{(\cdot)}(t)$ terms with corresponding confidence intervals to magnitude of influence and statistical significance for each causal variable. \newTxtR{More details are provided in Sections 3 and 4 of the Supporting Information.}

\begin{figure}[!t]
    \centering
    
    \includegraphics[trim={0 0 0 0mm}, clip, width=\textwidth]{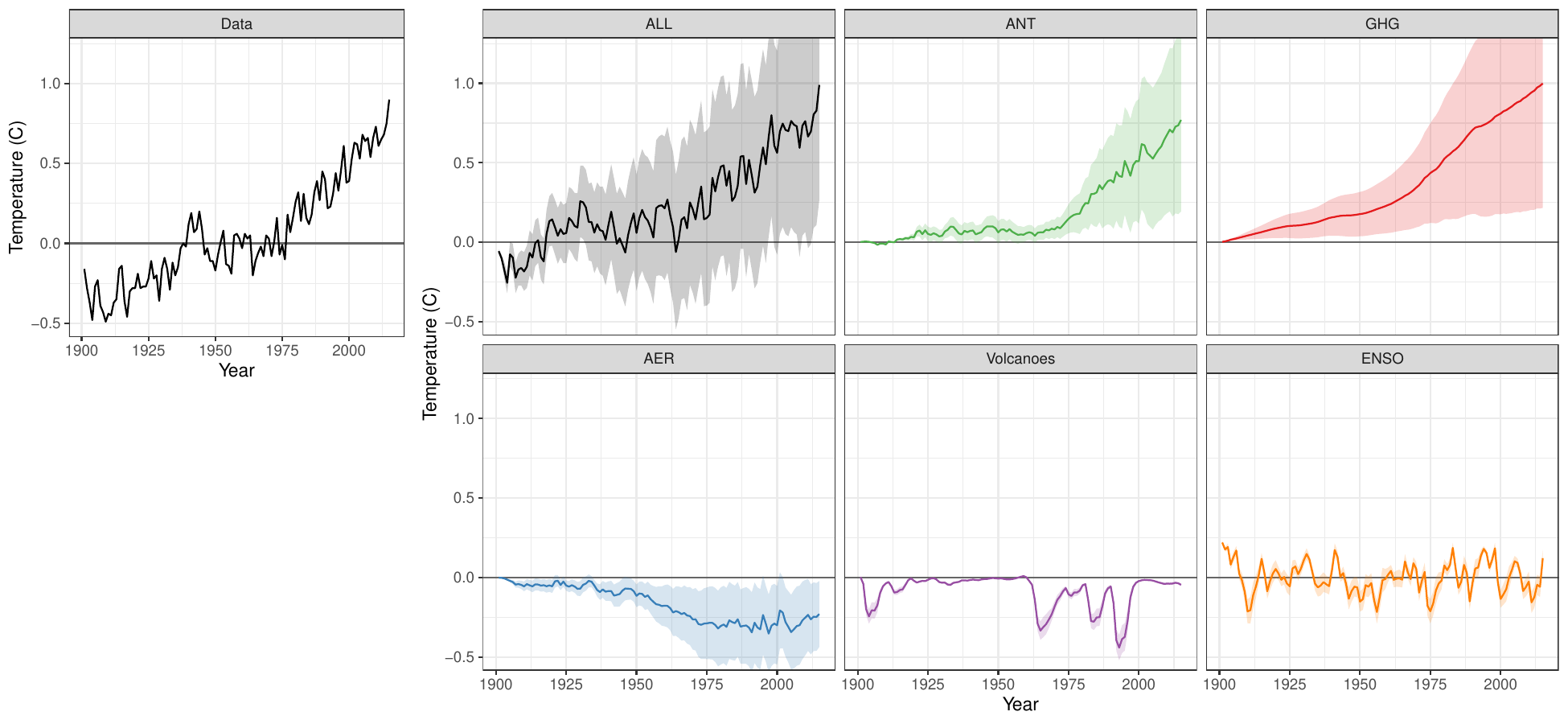}
    \caption{Reconstructing the effect of each covariate on the ``original'' scale, i.e., cumulatively summing the first-order differences. Trajectories include a \st{90\%} \newTxtR{``likely'' (66\%)} confidence interval. ``ALL" refers to all covariates, ``ANT'' is the combined anthropogenic radiative forcing, ``GHG'' and ``AER'' are the greenhouse gas and anthropogenic aerosol contributions (respectively; recall ANT = GHG + AER), ``Volcanoes'' represents the effect of volcanic stratospheric aerosol optical depth, and ``ENSO'' shows the effect of the El Ni\~no/Southern Oscillation.}
    \label{fig_cont}
\end{figure}

However, in this example where nonstationarity of the response and covariates required analysis of first-order differences, it's not immediately clear the extent to which statements about the $m^*_{(\cdot)}(t)$ are meaningful. Instead, note that we can reconstruct the original data from the increments as follows: for example, since $Y^*(t) = Y(t)-Y(t-1)$, $Y(t) = \sum_{i=1}^t Y^*(i)$.
Similarly, \newTxtR{we can reconstruct the variable-specific effects as $m_{(\cdot)}(t) = \sum_{i=1}^t m_{(\cdot)}^*(i)$.} \st{reconstructions can be made for the $m^*_{(\cdot)}(t)$ terms} \newTxtR{Since this is a linear transformation of the $m^*_{(\cdot)}(t)$, we can obtain corresponding estimates of their uncertainities; see Section 3 of the Supporting Information for further details}. \st{Since this reconstruction is a linear transformation of the first-order differences, it is straightforward to translate uncertainties for the coefficients $\{\mu, \alpha_i, \beta_i, \gamma_{\text{Volc},i}, \gamma_{\text{ENSO},i}\}$ into uncertainties for the $m_{(\cdot)}(t) = \sum_{i=1}^t m^*_{(\cdot)}(i)$ terms.}
Results are shown in Figure~\ref{fig_cont}, where we re-plot the original $Y(t)$ data for comparison. Here, ``ALL" refers to all covariates \newTxtR{(i.e., reconstructing the data from the linear model)}, ``ANT'' is the combined anthropogenic radiative forcing, ``GHG'' and ``AER'' are the greenhouse gas and anthropogenic aerosol contributions (respectively), ``Volcanoes'' represents the effect of volcanic stratospheric aerosol optical depth, and ``ENSO'' shows the effect of the El Ni\~no/Southern Oscillation. \newTxtR{Each trajectory includes a ``likely'' (66\%) confidence interval \citep[borrowing language from][]{mastrandrea2010guidance}.} As we would expect, greenhouse gas forcing drives increases to global temperatures, while anthropogenic aerosol and volcanic aerosols cause decreases to global surface temperatures. The El Ni\~no/Southern Oscillation drives year-to-year variability in global surface temperatures that correspond to El Ni\~no and La Ni\~na cycles.

\color{black}
Trend attribution seeks to determine whether radiative forcing has significantly altered the present-day GMST anomalies, relative to the pre-industrial climate, above and beyond any natural (non-anthropogenic) variations in the Earth system. The statistical counterfactual approach can answer this question by  defining two climate ``scenarios'':

\begin{enumerate}
    \item {Pre-industrial (PI)}: year-1901 radiative forcing and years for which the effect of both ENSO and volcanoes was zero (identified as 2003 and 1957, respectively).
    \item {Present-day (PD)}: 2015 levels of anthropogenic radiative forcing and the same years for ENSO and volcanoes.
\end{enumerate}
These scenarios are counterfactual in the sense that neither correspond to conditions that actually occurred in Nature. Furthermore, note that the autoregressive effect of each covariate is already encoded in estimates of the first-order differences $m^*_{(\cdot)}(t)$ and hence the cumulatively-summed $m_{(\cdot)}(t)$ terms shown in Figure~\ref{fig_cont}.
We are then interested in the expected change in GMST for the PD scenario relative to the PI scenario: $\Delta_\text{ANT} = m(\text{PD}) - m(\text{PI})$, where

\begin{equation} \label{eq:scen}
\begin{array}{l}
    {m}(\text{PI}) = \mu + m_X(1900) + m_{Z_\text{volc}}(1957) + m_{Z_\text{ENSO}}(2003) \\[1ex]
    {m}(\text{PD})  = \mu + m_X(2015) + m_{Z_\text{volc}}(1957) + m_{Z_\text{ENSO}}(2003),
\end{array}
\end{equation}
such that $\Delta_\text{ANT} = m_X(2015) - m_X(1900)$.
The Granger-causal tests conducted in Section~\ref{sec_example_sub1} already showed that we can reject the null hypothesis of no change, i.e., $H_0: \Delta_\text{ANT} = 0$; all that is left to do is to obtain best estimates and confidence intervals for the difference. Results are shown in Table~\ref{tab:example}: our best estimate is that anthropogenic forcing causes an increase of $0.77^\circ$C in global mean surface temperature, with a 90\% confidence of $(0.19^\circ\text{C}, 1.34^\circ\text{C})$. 
\color{black}
This result is consistent with conclusions from the  Intergovernmental Panel on Climate Change Sixth Assessment Report \citep[AR6;][]{IntergovernmentalPanelonClimateChangeIPCC2023}, which found that ``the likely range of total human-caused global surface temperature increase from 1850–1900 to 2010–2019 is $0.8^\circ\text{C}$ to $1.3^\circ\text{C}$, with a best estimate of $1.07^\circ\text{C}$.'' 
\newTxt{We note that the IPCC statements are based on analysis of GCM output using Pearl causality.}
The fact that our Granger-based conclusions are in alignment with the Pearl-causal analyses summarized by the IPCC AR6 both verifies that the statistical counterfactual is getting the right answer for the right reason while increasing our overall confidence in the human influence on GMST.

\begin{table}[!t]
\centering
\small
\caption{Best estimates, ``likely'' (66\%) confidence intervals, and $P$-values for the GMST change $\Delta$ due to anthropogenic forcing as well as ENSO and volcanic aerosols. $P$-values correspond to a the null hypothesis that each quantity is zero. \newTxtR{All results are based on analysis of lagged first-order differences, since the original measurements (without taking first-order differences) fail a formal test for stationarity.} \st{We show the results of analyzing lagged first-order differences as well as the results of a multiple regression without lags of the original measurements (i.e., without taking first-order differences). Results for the original measurements are shown only for comparison and are not discussed in this section.}}
\vskip2ex
\begin{tabular}{l|rrr}
\textbf{Quantity} & \textbf{Best estimate} & \textbf{``Likely'' Confidence Interval} & \textbf{$P$-value}  \\ 
\hline 
$\Delta_\text{ANT}$ & $0.77^\circ$C & $(0.19^\circ\text{C}, 1.34^\circ\text{C})$ & $0.07$ \\
$\Delta_\text{GHG}$ & $1.00^\circ$C  & $(0.21^\circ\text{C}, 1.78^\circ\text{C})$ & -- \\
$\Delta_\text{AER}$ & $-0.35^\circ$C  & $(-0.59^\circ\text{C}, -0.11^\circ\text{C})$ & -- \\ 
\hline 
$\Delta_\text{Volc}$ & $-0.44^\circ$C  & $(-0.52^\circ\text{C}, -0.36^\circ\text{C})$ & $<0.001$   \\
$\Delta_\text{ENSO}$ & $0.43^\circ$C  & $(0.37^\circ\text{C}, 0.50^\circ\text{C})$ & $<0.001$  \\
\end{tabular}
\label{tab:example}
\end{table}

The statistical counterfactual can similarly be used to extract estimates of change due to forcing- and driver-specific factors. 
For each variable, similar to the above example for the sum-total anthropogenic radiative forcing, we can identify years with the maximum and minimum effect in the corresponding $m_{(\cdot)}(t)$ while setting all other variables to years in which their covariates are close to zero. This procedure also applies to the two components of anthropogenic forcing, such that we can isolate the GHG and AER components separately (note, however, that these two forcing agents have the same regression coefficients, the $\{\beta_i\}$). 
\color{black}
As with $\Delta_\text{ANT}$, all of these change estimates are simply a rescaling of the associated coefficients but are more interpretable since they combine the autoregressive terms and have units matching those of GMST, which also allows us to inter-compare the relative influence of, e.g., volcanic activity and ENSO.

These results are also shown in Table~\ref{tab:example}: we estimate that GHG forcing contributed to an increase in GMST of between $0.21^\circ\text{C}$ and $1.78^\circ\text{C}$; anthropogenic aerosol forcing contributed to a decrease in GMST of between $0.11^\circ\text{C}$ and $0.59^\circ\text{C}$; volcanic forcing contributed to a decrease in GMST of between $0.36^\circ\text{C}$ and $0.52^\circ\text{C}$; and ENSO drives changes in GMST between $0.37^\circ\text{C}$ and $0.50^\circ\text{C}$. Once again, these estimates are consistent with Pearl-causal IPCC conclusions \citep{IntergovernmentalPanelonClimateChangeIPCC2023}: the AR6 states that GHG forcing contributed to a warming of $1.0^\circ\text{C}$ to $2.0^\circ\text{C}$, other human drivers (primarily aerosols) contributed to a cooling of $0^\circ\text{C}$ to $0.8^\circ\text{C}$, natural (volcanic and solar) forcing changed GMST by $-0.1^\circ\text{C}$ to $0.1^\circ\text{C}$, and internal variability changed GMST by $-0.2^\circ\text{C}$ to $0.2^\circ\text{C}$. 
This agreement increases confidence in Granger causal attribution and our statistical counterfactual methodology. 
\color{black}

\subsection{\newTxtR{Quantifying and interpreting causality: extreme event attribution}} \label{subsec:interp:ea}

\newTxtR{Finally, we demonstrate how the statistical counterfactual approach can be used to determine the anthropogenic influence on extreme events, as is commonly done for extreme event attribution \citep{national2016attribution}. Suppose the extreme event of interest is experiencing a GMST anomaly at least as large as the all-time maximum from the period we analyze (1900-2015): an anomaly of $0.9^\circ$C above the 1950-1980 climatology, which occurred in 2015. To answer this question, the statistical counterfactual can be used to obtain the ``factual'' distribution of GMST in 2015, the year in which the record occurred, as well as the counterfactual distribution of GMST for an arbitrary combination of inputs. The Gaussian assumptions underlying the unrestricted model implies that the factual and counterfactual GMST distributions will also be Gaussian; their means can be obtained as in Equation~\ref{eq:scen}, and their corresponding uncertainties can be obtained following the techniques described in Section 3 of the Supporting Information.}

\newTxtR{Here, we define four difference climate ``scenarios'':}
\begin{enumerate}
    \item[S1.] 1901 anthropogenic forcing with the 2015 ENSO state and volcanic activity (a counterfactual distribution).
    \item[S2.] 1990 anthropogenic forcing with the 2015 ENSO state and volcanic activity (a counterfactual distribution).
    \item[S3.] 2015 anthropogenic forcing with the 2015 ENSO state and volcanic activity (the ``factual'' distribution).
    \item[S4.] 2015 greenhouse gas forcing (zeroing out the aerosol contribution) with the 2015 ENSO state and volcanic activity (a counterfactual distribution).
\end{enumerate}
\newTxtR{In all cases, we set the ``background'' conditions (ENSO and volcanic aerosols) to their actual values from 2015, the year in which the event occurred. The scenarios otherwise evaluate a range of anthropogenic forcing values to assess its evolving effect on GMST. As described above, each scenario has a specific mean and standard deviation that defines a Gaussian distribution. The probability of experiencing a GMST at least as large as $0.9^\circ$C in each scenario is}
\[
P_{j} = 1-\Phi_j(0.9^\circ\text{C}), \hskip4ex j=1,\dots, 4,
\]
\newTxtR{where $\Phi_j(\cdot)$ is the Gaussian cumulative distribution function for scenario $j$. To compare these probabilities, we define three different probability ratios or ``risk ratios'' \citep{Paciorek2018quantifying}:}
\begin{equation}
    RR_1 = \frac{P_3}{P_1}, \hskip2ex RR_2 = \frac{P_3}{P_2}, \hskip2ex \text{and} \hskip2ex RR_3  =  \frac{P_4}{P_3}.
\end{equation}
\newTxtR{The first risk ratio compares the present-day and pre-industrial probabilities of experiencing a GMST anomaly of $\geq0.9^\circ$C; the second compares the present-day with the recent past; and the third assesses how much more likely the GMST anomaly might have been without the offsetting cooling from anthropogenic aerosols. Uncertainties in the probabilities $\{P_j: j=1, \dots, 4\}$ and risk ratios $\{RR_k: k = 1, \dots, 3\}$ are obtained via bootstrapping.}

\begin{figure}[!t]
    \centering
    \includegraphics[trim={0 0 0 0mm}, clip, width=0.8\textwidth]{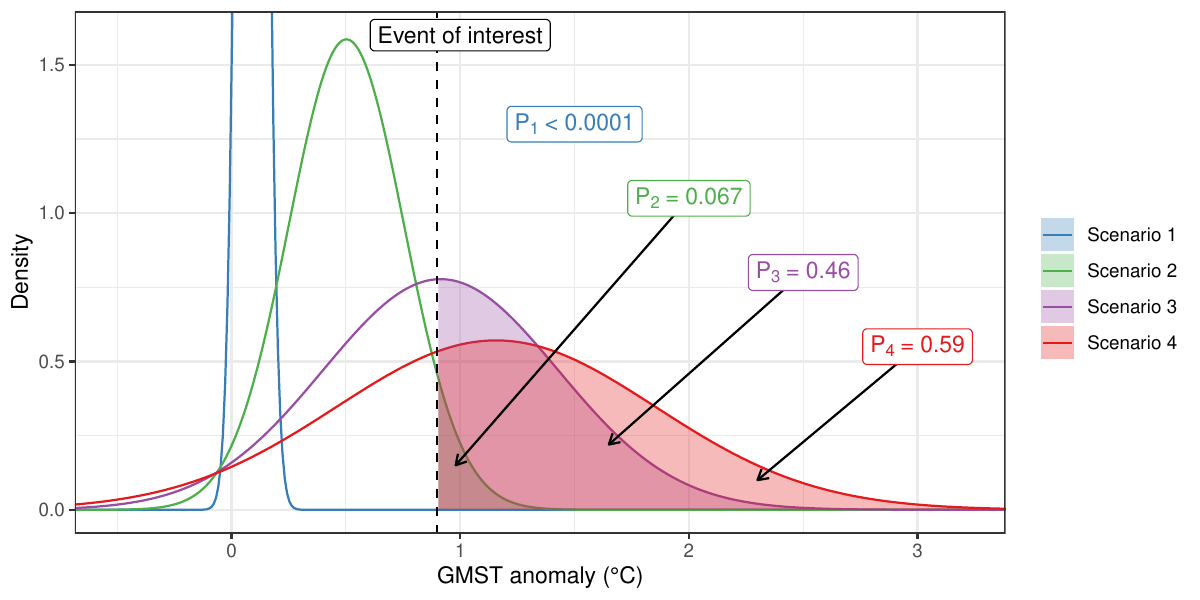}
    \caption{Scenario-specific GMST distributions for the three counterfactual (scenarios 1, 2, and 4) and one ``factual'' (scenario 3) climate scenarios. The vertical dashed line shows the event of interest: a GMST anomaly of $0.9^\circ$C. Also shown in shading (and labels) are the probabilities $\{P_j:j=1,\dots,4\}$ of experiencing a GMST anomaly of at least $0.9^\circ$C in each scenario.}
    \label{fig_eea}
\end{figure}

\begin{table}[!t]
\centering
\small
\caption{Best estimates and ``likely'' (66\%) confidence intervals for the three risk ratios comparing probabilities of experiencing a GMST anomaly of at least $0.9^\circ$C under four different climate scenarios.}
\begin{tabular}{ccc}
\textbf{Quantity} & \textbf{Best estimate} & \textbf{``Likely'' Confidence Interval}  \\ 
\hline \noalign{\smallskip}
$RR_1$ & $1.3\times10^{49}$ & $(2.3\times 10^{49}, 9.9 \times 10^{68})$ \\
$RR_2$ & $6.95$ & $(2.60, 4.42)$ \\
$RR_3$ & $1.27$ & $(1.07, 1.71)$ 
\end{tabular}
\label{tab:eventattr}
\end{table}

Event attribution results are shown in Figure~\ref{fig_eea} and Table~\ref{tab:eventattr}. In Figure~\ref{fig_eea}, we see that the event of interest is effectively impossible (i.e., probability $<0.001$) in a pre-industrial climate (Scenario 1) with 1901 levels of anthropogenic forcing. The event is still somewhat unlikely even in the recent past (1990 levels of anthropogenic forcing), with a probability of $0.067$ in Scenario 2. Unsurprisingly, the event is actually quite common in the present-day (with probability $0.46$ in Scenario 3), and the event would have been even more likely (probability $0.59$) without the offsetting cooling from anthropogenic aerosols (Scenario 4). Table~\ref{tab:eventattr} formally compares the ratios of the different probabilities along with their uncertainties. Again, we see that $RR_1$ is effectively $\infty$ since the denominator (the pre-industrial probability) is essentially zero: in other words, an anomaly of $>0.9^\circ$C is ``impossible'' without present-day anthropogenic forcing. Interestingly, from $RR_2$ we see that a GMST anomaly of $>0.9^\circ$C is nearly seven times more likely (and at least 2.6 times more probable) with 2015 levels of anthropogenic forcing compared to the recent past. Finally, from $RR_3$ we see that a GMST anomaly of $>0.9^\circ$C in 2015 would have been $27\%$ more likely (at least $7\%$ more likely) without the offsetting effects of anthropogenic aerosol forcing.
\color{black}

\section{A path forward: hybrid Granger and Pearl methods} \label{sec:hybrid}

As discussed in Section~\ref{sec:intro}, Pearl- and Granger-causal methods have both strengths and weaknesses. On one hand, Pearl methods are based on experimental intervention and counterfactual theory and therefore establish clear causal statements. On the other hand, for climate change D\&A, Pearl statements rely on climate models, which means Pearl attribution studies may require high computational costs associated with running physical and dynamical models. Furthermore, attribution conclusions based on GCMs are subject to structural uncertainties and are only ``reliable'' insofar as the models are fit-for-purpose for simulating the variables and spatial scales of interest. Granger-causal attribution conclusions are not as controlled and are hence weaker than Pearl-causal statements, but their primary reliance on observations from Nature means that (1) one can conduct attribution statements with very low computational cost and in near-real time, and (2) there is no concern as to whether the input data are ``representative'' of the real world (as is the case with GCMs). 
\newTxt{More generally, GC methods are likely to be of most use where climate models are expected to perform poorly due to unrealized or poorly represented processes.}



A natural question then arises: to what extent can Granger and Pearl causality be combined to leverage the strengths and minimize the weaknesses of both methods? \st{One example of a recent study that combined Pearl and Granger causal methods considered climate change attribution for regional precipitation change and is described in a pair of papers (Risser et al., 2022, 2024).}
Despite the importance of understanding the magnitude of human-induced changes to mean and extreme precipitation at localized scales, robust conclusions for sub-continental changes remain difficult to obtain. Many attempts to attribute local-scale trends in rainfall are largely inconclusive \citep{knutson2018model, kirchmeier2020human, Huang2021Northeast, christidis2022human}, even for places such as the United States where there are well-documented century-length trends in mean and extreme precipitation \citep{Kunkel2003,Easterling2017,risser2019detected}.
\st{One of the biggest reason for low confidence in local attribution is model uncertainty (Sarojini et al., 2016), which}
\newTxtR{As described in \cite{Sarojini2016detection}, there are three primary reasons for low confidence in regional attribution statements for precipitation: observational uncertainty, large internal variability, and modeling (or structural) uncertainty. Model uncertainty in particular}
is exacerbated for precipitation attribution since a treatment of anthropogenic aerosols is critical \citep{hegerl2015challenges}. 
\sti{Regarding aerosols, GCMs have two challenges: first, there is uncertainty regarding how aerosols are represented within the model (e.g., prescribed versus parameterized), and second, the precipitation response to aerosols is highly uncertain.}

Given the limitations of GCMs with respect to simulating regional precipitation change, it is natural to pivot from model-based Pearl causality to observations-based Granger causality. However, Granger causality for regional precipitation change is not straightforward: the low signal-to-noise ratio associated with present-day regional precipitation change, the complicated relationships between precipitation and anthropogenic aerosols, and the magnitude of internal variability make it challenging to propose an appropriate statistical counterfactual model as in Equation~\ref{eq:statCF}.
\newTxt{For example, what causal forcing agents should be included, and are there cross-correlations between different forcing agents? Are the relevant forcing agents separable? What are the relevant low-frequency drivers for inclusion as mediators, and do the external forcing agents influence their relationship with precipitation? Given the importance of anthropogenic aerosols, how can aerosol forcing be represented in an observational analysis? And, potentially most importantly, is an additive framework appropriate for precipitation?}

Starting with the continental United States (CONUS) and considering the recent historical record, \cite{Risser2022framework} and \cite{Risser2024anthropogenic} utilized a hybrid approach to D\&A by (1) using Pearl-causal methods and climate model simulations to identify an appropriate formula for modeling a time series of precipitation and then (2) setting climate models aside and applying the derived formula to in situ records from rain gauge measurements to generate Granger-causal attribution statements. 
\sti{We now briefly summarize the Pearl-causal tests conducted for the first step (for complete details, see Risser et al., 2022)}.
\sti{Let $P(t), t = 1, \dots, T$ represent annual summaries of mean or extreme precipitation from a given weather station. Risser et al. (2022) propose a very general formula for modeling $P(t)$:}
\sti{There are numerous questions regarding how best to specifically model each component of Equation 6. For example, what forcing agents should be included in $P_F(t)$, and are there cross-correlations between different forcing agents? Are the relevant forcing agents separable? What are the relevant low-frequency drivers, and do the external forcing agents influence their relationship with precipitation? Since anthropogenic aerosols are one of the relevant forcing agents, how can aerosol forcing be represented in an observational analysis? Finally, one must also verify that the additive framework in Equation 6 is appropriate for precipitation, as opposed to a multiplicative model.}
Each question from the previous paragraph was tested using output from tailored experiments with GCMs, i.e., using Pearl Causality. For example, to
\sti{one of the important questions raised above involves the specific external forcing agents that should be included in the forced component $P_F(\cdot)$ in Equation 6. In principle, a large number of variables could be included: greenhouse gas forcing, anthropogenic aerosols, stratospheric ozone, solar and volcanic forcing, and land-use/land-cover change. To }
quantitatively determine which causal forcing agents should be included, \cite{Risser2022framework} used single-forcing, coupled model experiments from the Detection and Attribution Model Intercomparison Project \citep[DAMIP;][]{gillett2016detection} and the Land Use Model Intercomparison Project \citep[LUMIP;][]{lawrence2016land}.
\sti{DAMIP and LUMIP provide single-forcing, coupled model experiments for each of the aforementioned forcing agents; for each experiment, Risser et al. (2022) calculate trends in seasonal mean and extreme precipitation separately for each General Circulation Model that participated in the experiments. Ultimately, looking across the calculated trends from the different models, for seasonal mean and extreme precipitation it was clear that one could safely ignore the influence of land-use/land-cover change, stratospheric ozone, and solar/volcanic forcing; on the other hand, it was similarly clear that it was critical to account for greenhouse gas and anthropogenic aerosol forcing (see Figure 4 of Risser et al., 2022).}
\newTxt{From these experiments, high-level decisions regarding the relevant causal factors for inclusion in the unrestricted model were made.}
\sti{The various hypotheses considered and conclusions from each test are summarized in Table 3 of Risser et al. (2022). Importantly, all tests were conducted separately for each GCM with no attempts made to aggregate output from different models: in this way, structural uncertainties were obviated. The}
\newTxt{The suite of}
tests \citep[see Table 3 of][]{Risser2022framework} resulted in a simplified autoregressive model that could be estimated using century-length records of gauged measurements; see Equation 2 in \cite{Risser2024anthropogenic}. 
\sti{The authors include the caveat that all conclusions derived in Risser et al. (2022) about the appropriateness of the D\&A formula only apply to rain gauge measurements from the CONUS from the last 120 years. There is no guarantee that the conclusions would hold for other regions of the globe or in the future under increased anthropogenic global warming. }
Ultimately, \cite{Risser2024anthropogenic} used the statistical counterfactual approach, i.e., Granger causality, to generate a conclusive attribution statement about the role of anthropogenic climate change on seasonal changes to mean and extreme precipitation in the United States.
\newTxtR{Note that the response variables $Y$ and a large majority of the covariates ($X$ and $Z$) used in \cite{Risser2024anthropogenic} pass an augmented Dickey-Fuller test for stationarity (see Supplemental Table S2), and hence reasonably meet the requirements for Granger causality discussed previously.}
Since the underlying formula for conducting Granger-based attribution was vetted using Pearl-causal methods applied to climate models, \cite{Risser2024anthropogenic} avoid at least some of the usual limitations associated with Granger-causal statements, strengthening the resulting attribution conclusions.

Of course, the approach outlined in \cite{Risser2022framework} and \citep{Risser2024anthropogenic} is just one possible way to combine Pearl and Granger perspectives on climate change attribution. Clearly, both perspectives on D\&A are valuable, and we by no means advocate for one over the other. In tandem, Granger and Pearl causality provide two benefits: first, confidence is enhanced when both methods yield the same conclusions; and second, there are numerous examples of rapid Granger-causal D\&A providing motivation for more in-depth (and more computationally intensive) Pearl-causal studies. For example, the results from \cite{Risser2017} regarding the anthropogenic influence on the extreme rainfall associated with Hurricane Harvey provided the motivation for \cite{Patricola2018anthropogenic}, which used storyline/pseudo-global warming dynamical experiments to make a broader statement on the human influence on rainfall associated with tropical cyclones. We anticipate that continued efforts to combine these two paradigms for D\&A will advance the scientific community's understanding of if, how, and why anthropogenic activities impact different aspects of the global climate system. 


%
\section{Discussion} \label{sec:conclusions}

\newTxt{In this work, we have presented a formal definition of Granger causality and how it can be used for climate change D\&A while providing a clear statement of the implied causal framing. We furthermore describe the so-called \textit{statistical counterfactual} approach for quantifying and interpreting a Granger-causal attribution statement. The proposed methodology is demonstrated on a widely-studied attribution problem, and we show how Granger attribution statements agree with existing Pearl-causal results, ultimately strengthening the overall evidence for the human influence on global mean surface air temperature changes. Finally, we argue for hybrid approaches to climate change D\&A that use leverage the strengths of both Pearl and Granger causality while minimizing at least some of their limitations.}

\newTxt{One of the main benefits of using Granger causality methods is its reliance only on observations, which avoids the structural (or model) uncertainty associated with climate models. However, in some cases, observations required for the Granger approach (e.g., global radiative forcing from aerosols, as in Section~\ref{sec_sc_eg}) are highly uncertain and might vary across different data products. In this case, it becomes extremely important to account for uncertainties in the $X$ and or $Z$ variables using so-called ``error-in-variable'' methods. We refer the interested reader to \cite{Lau2023extreme} and \cite{Risser2024anthropogenic} for two approaches for dealing with this issue. We also note that input uncertainties must be accounted for in Pearl-causal methods like optimal fingerprinting \citep[see, e.g.,][]{Katzfuss2017bayesian}.
}

\newTxt{As mentioned in Section~\ref{sec:intro}, one of the most public-facing attribution efforts is the World Weather Attribution (WWA) project. The WWA project uses both observations and GCMs to conduct rapid attribution in the aftermath of an extreme weather event. Given its high visibility and connections to the methods described in Section~\ref{sec:hybrid}, we briefly compare and contrast the two approaches to climate change D\&A.
The WWA methodology \citep{ascmo-6-177-2020} involves an eight-step protocol; most relevant to this discussion are the ``trend detection'' and ``multi-method, multi-model attribution'' steps. The trend detection step analyzes observational data to determine whether there is a detectable trend above and beyond natural variability for the event of interest. Assuming that a trend is present (and after a careful assessment of climate models), attribution conclusions are subsequently derived using either (1) direct comparison of a pair of fixed forcing runs, or (2) an analysis of transient forcing runs that is identical to the observational analysis. The specific statistical model used in the trend detection (and analysis of transient runs) mirrors the unrestricted model in Equation~\ref{eq:GCunrestr}, wherein careful choices are made about the form of the error term and which confounding variables to include (however, the WWA methodology generally does not include autoregression). In this sense, the hybrid approach to D\&A described in Section~\ref{sec:hybrid} can be seen as an ``inversion'' of the WWA approach to extreme event attribution: using the climate models (not observations) to build the statistical model and the observations (not models) to attribute. Interestingly, the WWA methodology paper \citep{ascmo-6-177-2020} does not indicate how the attribution statements should be interpreted from a causal standpoint.} \st{From our perspective, we note that the choice of GCM experiments for attribution in the WWA approach determines the causal framing: while Pearl causality applies to the analysis of a pair of fixed forcing, we argue that analyzing transient runs implies Granger causality even though it uses output from dynamical models.}

\newTxt{Whichever approach is taken (using observations to build a model and  dynamical model output for attribution, or vice versa), it is important to keep in mind the risk of drawing misleading attribution conclusions from a mis-specified statistical model. For example, if an important confounder or mediator is omitted, the ``true'' causal effect of  that variable may be projected onto the causal factor of interest, yielding a false attribution conclusion. For a given analysis, there is of course always a tradeoff to be made between including too many confounding variables (leading to overfitting the training data) and having enough degrees of freedom to infer the regression coefficients and distribution of the error term. Statistical learning techniques such as regularization \citep{tibshirani1996regression, james2013introduction} can help, although for climate change D\&A the problem is generally scientific, not statistical. We argue that the risk of these problems depends on the underlying scientific understanding of the mechanisms (both anthropogenic and natural) that drive the event class of interest. Ultimately, we reiterate the importance of using Pearl-causal studies with climate models in tandem with observationally-based Granger-causal conclusions, since their agreement or disagreement will always yield important insight into how anthropogenic activity impacts extreme weather and climate events.}

\newTxt{Lastly, there is a broader philosophical point to discuss when comparing experiment-based causation with observation-based causation. Consider the case where the observations exhibit a much larger climate change trend than the climate models: under the Pearl approach, only part of the observed trend would be considered attributable to climate change, with the remaining trend perhaps ascribed to factors not captured in the models (land use changes, feedback systems, or similar) or due to poor model performance (in which case the overall attribution statement might be revised under expert judgment). In this case, we risk understating the effects of climate change, and make a conservative attribution statement. Under the combined approach proposed in Section~\ref{sec:hybrid}, if we used the same climate models to determine the relevant factors (although of course, this should be done after careful consultation with domain experts), we would be likely to attribute more, and maybe even all, of the observed trend to climate change. However, if the climate models still lack some real-world processes (deforestation, say), we risk overstating the effects of climate change. Ultimately, the choice of which method to use depends on risk preferences \citep{Lloyd2018}, which should always inform methodological decisions about a given climate change D\&A study.}

\color{black}
\section*{Acknowledgments}

This research was supported by the Director, Office of Science, Office of Biological and Environmental Research of the U.S. Department of Energy under Contract No. DE-AC02-05CH11231 and by the Regional and Global Model Analysis Program area within the Earth and Environmental Systems Modeling Program %
as part of the Calibrated and Systematic Characterization, Attribution, and Detection of Extremes (CASCADE) project. 


This document was prepared as an account of work sponsored by the United States Government. While this document is believed to contain correct information, neither the United States Government nor any agency thereof, nor the Regents of the University of California, nor any of their employees, makes any warranty, express or implied, or assumes any legal responsibility for the accuracy, completeness, or usefulness of any information, apparatus, product, or process disclosed, or represents that its use would not infringe privately owned rights. Reference herein to any specific commercial product, process, or service by its trade name, trademark, manufacturer, or otherwise, does not necessarily constitute or imply its endorsement, recommendation, or favoring by the United States Government or any agency thereof, or the Regents of the University of California. The views and opinions of authors expressed herein do not necessarily state or reflect those of the United States Government or any agency thereof or the Regents of the University of California.

\bibliography{references}

\clearpage


\begin{appendix}
\renewcommand{\thefigure}{S\arabic{figure}}
\renewcommand{\thetable}{S\arabic{table}}
\renewcommand{\theequation}{S\arabic{equation}}

\section{Supplemental Figures}

\begin{figure}[!h]
    \centering
    \color{black}
    \includegraphics[trim={0 0 0 0mm}, clip, width=\textwidth]{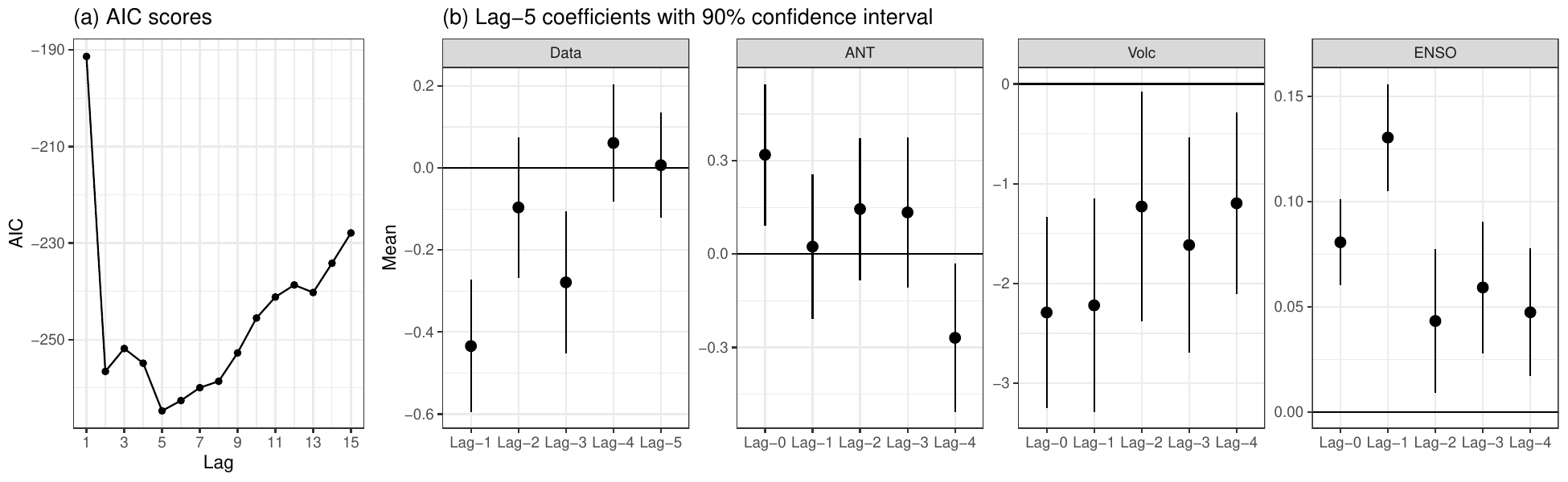}
    \caption{(a) Akiake's Information Criteria (AIC) for the unrestricted autoregressive model across a variety of time lags; $p=1$ corresponds to no autoregression. Smaller AIC indicates a better fit. (b) Lagged coefficient estimates from the unrestricted model with 90\% confidence intervals, using the best-performing statistical model according to the AIC.}
    \label{fig_aic}
\end{figure}

\begin{figure}[!h]
    \centering
    \color{black}
    \includegraphics[trim={0 0 0 0mm}, clip, width=\textwidth]{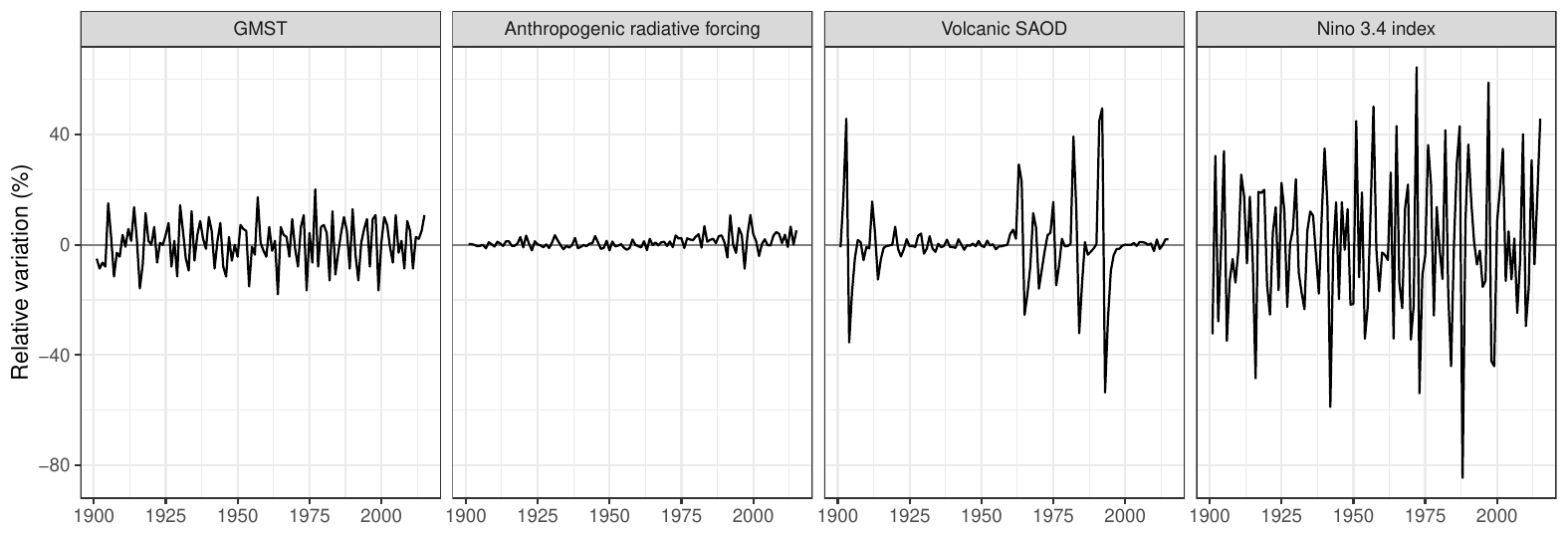}
    \caption{First-order differences for the three covariates used in the example in Section 4 of the main text. The first-order differences are normalized by the range of the original measurements (1.39 units, 2.19 units, 2.65 units, and 0.11 units for GMST, ANT, Nino 3.4, and vSAOD, respectively) in order to show their variation relative to the ``raw'' covariates.}
    \label{fig_relvar}
\end{figure}

\clearpage

\section{Supplemental Tables}

\begin{table}[!h]
\centering
\color{black}
\caption{Results of testing whether each variable of interest is stationary, and furthermore whether its first-order differences are stationary. The $P$-values shown in parentheses are from an augmented Dickey-Fuller test, where the null hypothesis is that the data are non-stationary (small $P$-values indicate this can be rejected in favor of the data being stationary).}
\vskip2ex
\begin{tabular}{lcc}
\hline\noalign{\smallskip}
\textbf{Variable} & \textbf{Stationary? ($P$)} & \textbf{1st-order diff. stationary? ($P$)} \\ 
\hline\noalign{\smallskip}
Global mean surf. temp. & No ($0.60$) & Yes ($<0.01$) \\
Anthro. radiative forcing & No ($>0.99$) & Yes ($<0.01$) \\
Volcanic aerosols & Yes ($0.014$) & Yes ($<0.01$) \\
ENSO (Ni\~no3.4 index) &Yes ($<0.01$) & Yes ($<0.01$) \\
\noalign{\smallskip}
\hline
\end{tabular}
\label{tab:stationary}
\end{table}

\begin{table}[!h]
\caption{Results of testing whether the variables used in Risser et al. (2024) are stationary across each three-month season. Tests for seasonal mean and maximum precipitation and the local SO$_2$ emissions are for United States-averaged time series (all other variables are spatially-invariant). The $P$-values shown in parentheses are from an augmented Dickey-Fuller test, where the null hypothesis is that the data are non-stationary (small $P$-values indicate this can be rejected in favor of the data being stationary).}
\label{tab:my_label}
\vskip2ex
\centering
\begin{tabular}{l|cccc}
\textbf{Variable} & \textbf{Winter} & \textbf{Spring} & \textbf{Summer} & \textbf{Autumn}\\
\hline
Seasonal mean daily precipitation & $<0.01$ & $<0.01$ & $<0.01$ & $<0.01$\\
Seasonal maximum daily precipitation & $0.047$ & $0.019$ & $<0.01$ & $<0.01$\\ \hline
Slow response (GHG and global aerosol forcing) & $>0.99$ & $>0.99$ & $>0.99$ & $>0.99$\\
Fast response (local SO$_2$ emissions) & $0.97$ & $0.97$ & $0.98$ & $0.97$\\
ENSO longitude index & $<0.01$ & $<0.01$ & $<0.01$ & $<0.01$\\
Pacific-N. American teleconnection & $<0.01$ & $<0.01$ & $<0.01$ & $<0.01$\\
Arctic Oscillation & $0.02$ & $<0.01$ & $<0.01$ & $<0.01$\\
North Atlantic Oscillation & $0.02$ & $<0.01$ & $<0.01$ & $<0.01$\\
Atlantic Multidecadal Oscillation & $0.52$ & $0.36$ & $0.36$ & $0.52$\\
\end{tabular}
\end{table}

\clearpage
\section{Inference for the unrestricted model}

\subsection{Writing the unrestricted model in matrix notation}

In the main text, we wrote the unrestricted model as
\[
Y^*(t) = \mu + m^*_{Y}(t) + m^*_{X}(t) + m^*_{Z_\text{volc}}(t) + m^*_{Z_\text{ENSO}}(t) + \varepsilon(t),
\]
where
\[
m^*_{Y}(t) = \sum_{i=1}^{p} \alpha_i Y^*(t-i), \hskip2ex m^*_{X}(t) = \sum_{i=0}^{p-1} \beta_i X^*(t-i), \hskip2ex \text{and} \hskip1ex m^*_{Z_{(\cdot)}}(t) = \sum_{i=0}^{p-1} \gamma_{(\cdot),i} Z^*_{(\cdot)}(t-i).
\]
Alternatively, we can re-write this in matrix notation:
\[
{\bf Y}^* = {\bf W} \boldsymbol{\theta} + \boldsymbol{\varepsilon},
\]
where:
\begin{itemize}
    \item ${\bf Y}^* = \left[Y^*(1), \dots, Y^*(n) \right]$ is a vector of length $n=115$ (for the years $t=1901,\dots, 2015$),

    \item ${\bf W}$ is a matrix with $n$ rows and $1+4p$ columns:
    \[
    {\bf W} = \left[{\bf 1}_n \> {\bf L}_Y \> {\bf L}_X \> {\bf L}_\text{Volc} \> {\bf L}_\text{ENSO} \right],
    \]
    where, e.g., ${\bf L}_Y$ is a $n\times p$ matrix where the columns consist of the $p$ lagged values of $Y$. Note that ${\bf L}_X = {\bf L}_\text{GHG} + {\bf L}_\text{AER}$, where ${\bf L}_\text{GHG}$ and ${\bf L}_\text{AER}$ are the lagged greenhouse gas and aerosol forcing, respectively.
    

    \item $\boldsymbol{\theta} = (\mu, \alpha_1, \dots, \alpha_p, \beta_1, \dots, \beta_p, \dots)$ is a vector of length $1+4p$, and

    \item $\boldsymbol{\varepsilon} = \left[\varepsilon(1), \dots, \varepsilon(n) \right]$ is a vector of length $n=115$.
\end{itemize}
Recall that we assumed each $\varepsilon(t) \stackrel{\text{iid}}{\sim} N(0, \sigma^2)$. The implication is that ${\bf Y}^*$ follows a $n$-dimensional multivariate normal distribution:
\[
{\bf Y}^* \sim N_n({\bf W}\boldsymbol{\theta}, \sigma^2 {\bf I}_n).
\]

\subsection{Linear model theory for inference on $\sigma^2$ and $\boldsymbol{\theta}$}

Denote  ${\bf H} = ({\bf W}^\top {\bf W})^{-1}$. Standard results from statistical theory provide the following:

\begin{itemize}
    \item The best estimate of the regression coefficients $\boldsymbol{\theta}$ is
    $
    \widehat{\boldsymbol{\theta}} = {\bf H} {\bf W}^\top {\bf Y}^*$, 
    where $\widehat{(\cdot)}$ denotes a statistical estimate.

    \item Similarly, the best estimate of the output of the linear model (i.e., the reconstruction of the data as specified by the linear relationship with ${\bf W}$) is
    $
    \widehat{\bf Y}^* = {\bf W} \widehat{\boldsymbol{\theta}} = {\bf W} {\bf H} {\bf W}^\top {\bf Y}^*$.

    \item The best estimate of $\sigma^2$ is $
    \widehat{\sigma}^2 = \frac{1}{n} \sum_t \big[ Y^*(t) - \widehat{Y}^*(t) \big]^2$.

    \item The variance-covariance matrix of the estimated regression coefficients is
    $\text{Var}(\widehat{\boldsymbol{\theta}}) = \widehat{\sigma}^2 {\bf H}$. Since ${\bf W}$ has $1+4p$ columns, this variance-covariance matrix is of size $(1+4p)\times (1+4p)$.
    

    \item Properties of the multivariate normal distribution thus imply that: $\widehat{\boldsymbol{\theta}} \sim N_n( {\bf H}{\bf W}^\top{\bf Y}^*, \sigma^2 
    {\bf H})$.
\end{itemize}

\noindent \textbf{In summary:} our best estimate of the regression coefficients is $\widehat{\boldsymbol{\theta}}$, with uncertainties described by $\text{Var}(\widehat{\boldsymbol{\theta}})$, and this estimate follows a multivariate normal distribution.

\subsection{Reconstructing estimates of $m_{(\cdot)}(t)$}

All of the $m^*_{(\cdot)}(t)$ terms (as well as the cumulatively summed $m_{(\cdot)}(t)$ terms) can be written as linear combinations of the regression coefficients $\widehat{\boldsymbol{\theta}}$, which allows us to obtain both their estimates and uncertainties. The important result is that, if a vector ${\bf x}$ has a multivariate normal distribution with mean ${\bf u}$ and covariance ${\bf C}$, the linear combination ${\bf A x}$ also has a normal distribution with mean ${\bf Au}$ and covariance ${\bf A C A^\top}$. In order to keep track of all cross-covariances (which become important when calculating uncertainties for arbitrary linear combinations, let ${\bf m}^*$ denote a concatenated vector of all of the $m_{(\cdot)}^*(t)$ terms, and similarly ${\bf m}$ denote a concatenated vector of all of the $m_{(\cdot)}(t)$ terms: for example,
\[
{\bf m}^* = \left[ {\bf m}^*_\text{Data}, {\bf m}^*_Y, {\bf m}^*_X, {\bf m}^*_\text{GHG}, {\bf m}^*_\text{AER}, {\bf m}^*_\text{Volc}, {\bf m}^*_\text{ENSO} \right].
\]
Here, each ${\bf m}^*_{(\cdot)} = (m^*_{(\cdot)}(1), \dots, m^*_{(\cdot)}(n))$. Define
\[
\begin{array}{c}
     {\bf W}_Y = \left[{\bf 0}_n \> {\bf L}_Y \> {\bf 0}_{n\times p} \> {\bf 0}_{n\times p} \> {\bf 0}_{n\times p} \right] \\
     {\bf W}_X = \left[{\bf 0}_n \> {\bf 0}_{n\times p} \> {\bf L}_X \> {\bf 0}_{n\times p} \> {\bf 0}_{n\times p} \right] \\
     {\bf W}_\text{GHG} = \left[{\bf 0}_n \> {\bf 0}_{n\times p} \> {\bf L}_\text{GHG}  \> {\bf 0}_{n\times p} \> {\bf 0}_{n\times p} \right] \\
     {\bf W}_\text{AER} = \left[{\bf 0}_n \> {\bf 0}_{n\times p} \> {\bf L}_\text{AER} \> {\bf 0}_{n\times p} \> {\bf 0}_{n\times p} \right] \\
     {\bf W}_\text{Volc} = \left[{\bf 0}_n \> {\bf 0}_{n\times p} \> {\bf 0}_{n\times p} \> {\bf L}_\text{Volc} \> {\bf 0}_{n\times p} \right] \\
     {\bf W}_\text{ENSO} = \left[{\bf 0}_n \> {\bf 0}_{n\times p} \> {\bf 0}_{n\times p} \> {\bf 0}_{n\times p} \> {\bf L}_\text{ENSO} \right],     
\end{array}
\]
so that, e.g., ${\bf m}^*_Y = {\bf W}_Y \widehat{\boldsymbol{\theta}}$. Similarly, we can obtain ${\bf m}^* = {\bf W}_\text{stack} \widehat{\boldsymbol{\theta}}$, where ${\bf W}_\text{stack}$ is a $7n \times (1+4p)$ with the various ${\bf W}_{(\cdot)}$ matrices stacked on top of one another:
\[
{\bf W}_\text{stack} = \left[ \begin{array}{c}
     {\bf W}  \\
     {\bf W}_Y \\
     {\bf W}_X \\
     {\bf W}_\text{GHG} \\
     {\bf W}_\text{AER} \\
     {\bf W}_\text{Volc} \\
     {\bf W}_\text{ENSO} 
\end{array} \right].
\]
To convert back to the original scale from first-order differences, define a square, $n\times n$ lower triangular matrix with all ones on the diagonal and lower triangle:
\[
{\bf J} = \left[ \begin{array}{ccccc}
     1 & 0 & 0 & \cdots & 0 \\
     1 & 1 & 0 & \cdots & 0 \\
     1 & 1 & 1 & \cdots & 0 \\
     \vdots & \vdots & \vdots & \ddots & \vdots \\
     1 & 1 & 1 & \cdots & 1 
\end{array} \right].
\]
Then, ${\bf m} = ({\bf I}_7 \otimes {\bf J}) {\bf W}_\text{stack} \widehat{\boldsymbol{\theta}}$ and, following the rules above, 
\[
\text{Var}({\bf m}) = \sigma^2 ({\bf I}_7 \otimes {\bf J}) {\bf W}_\text{stack} {\bf H} {\bf W}_\text{stack}^\top ({\bf I}_7 \otimes {\bf J})^\top.
\]
As with other quantities, the original Gaussian assumptions imply that ${\bf m}$ also follows a Gaussian distribution.

The best estimates and uncertainties for the $\Delta$ terms in trend attribution (Section 4.2) and the scenario-specific normal distributions for event attribution (Section 4.3) can be obtained by picking out the appropriate components of ${\bf m}$ and $\text{Var}({\bf m})$ and applying the appropriate linear combination to these terms.

\clearpage
\section{Reproducibility code}

The following \texttt{R} code is used to generate all results in the main text.

\small
\begin{verbatim}
#==============================================================================
# Example: changes to global surface air temperature
# Section 4 of "Granger causal inference for climate change attribution"
# Mark Risser
# Lawrence Berkeley National Laboratory
# April, 2025
#==============================================================================

library(ggplot2)
library(gridExtra)
library(RColorBrewer)
library(tseries)
library(gridExtra)

#==============================================================================
# Load data
#==============================================================================
start.year <- 1900
end.year <- 2015
data_df <- read.csv("granger_example_data.csv")

#==============================================================================
# Plot data and covariates
#==============================================================================
# For nice plotting
plot_df <- data_df
plot_df$Nino34_ng1 <- data_df$Nino34
plot_df$Nino34_ng1[data_df$Nino34 > 0] <- 0
plot_df$Nino34_ng2 <- 0
plot_df$Nino34_ps1 <- data_df$Nino34
plot_df$Nino34_ps2 <- 0
plot_df$Nino34_ps1[data_df$Nino34 < 0] <- 0

# Combine anthropogenic variables
ant_long <- rbind(data_df, data_df, data_df)
ant_long$y <- c(data_df$GHG, data_df$AER, data_df$ANT)
ant_long$type <- factor(rep(c("GHG", "AER", "ANT = GHG + AER"), each = nrow(data_df)), 
                        levels = c("GHG", "AER", "ANT = GHG + AER"))

# Generate figure
pdf("Figure_data_covariates.pdf", width = 10.5, height = 6)
grid.arrange(
  ggplot(plot_df, aes(x = Year, y = GMST)) + geom_hline(yintercept = 0, color = "gray50") +
    geom_line() + theme_bw() + ggtitle("(a) Global mean surface temperature (rel. to 1950-1980)") +
    scale_x_continuous(breaks = c(seq(from=1900,to=2000,by=25), 2015)) +
    labs(x = NULL, y = "Temperature anomalies (°C)"),
  ggplot(ant_long, aes(x = Year, y = y, color = type)) + 
    geom_hline(yintercept = 0, color = "gray50") + geom_line() + theme_bw() + 
    ggtitle("(b) Anthropogenic radiative forcing (relative to 1900)") +
    scale_color_manual(values = c(brewer.pal(3,"Set1")[1:2], "black"), name = NULL) +
    scale_x_continuous(breaks = c(seq(from=1900,to=2000,by=25), 2015)) +
    labs(x = NULL, y = "Radiative forcing (W m^-2)") +
    theme(legend.position = c(0.2, 0.8)),
  ggplot(data_df, aes(x = Year, y = vSAOD)) + geom_hline(yintercept = 0, color = "gray50") +
    geom_ribbon(aes(ymax = vSAOD), ymin = 0, fill = "gray80") +
    geom_line() + theme_bw() + ggtitle("(c) Volcanic stratospheric aerosol optical depth") +
    scale_x_continuous(breaks = c(seq(from=1900,to=2000,by=25), 2015)) +
    labs(x = NULL, y = "SAOD at 550nm"),
  ggplot(data_df, aes(x = Year, y = Nino34)) + geom_hline(yintercept = 0, color = "gray50") +
    geom_ribbon(aes(ymin = Nino34_ng1, ymax = Nino34_ng2), fill = brewer.pal(3,"Set1")[2], alpha = 0.6) +
    geom_ribbon(aes(ymin = Nino34_ps1, ymax = Nino34_ps2), fill = brewer.pal(3,"Set1")[1], alpha = 0.6) +
    geom_line() + theme_bw() + ggtitle("(d) Nino 3.4 index") +
    scale_x_continuous(breaks = c(seq(from=1900,to=2000,by=25), 2015)) +
    labs(x = NULL, y = "SST anomalies (°C)"),
  nrow = 2
)
dev.off()

#==============================================================================
# Test for stationarity of original measurements
#==============================================================================
adf.test(data_df$GMST)
adf.test(data_df$ANT)
adf.test(data_df$vSAOD)
adf.test(data_df$Nino34)
# GMST and ANT are nonstationary!

#==============================================================================
# Calculate first-order differences
#==============================================================================
f_create_lagmat <- function(x, p){
  N <- length(x)
  mat <- matrix(0, N, p)
  for(i in 1:N){
    if(i <= p){
      if(i == 1){
        mat[i,p] <- x[i]
      } else{
        mat[i,(p-i+1):p] <- x[1:i]
      }
    } else{
      mat[i,] <- x[(i-p+1):i]
    }
  }
  return(mat)
}
data_diffs <- data_df[-1,]
for(i in 2:7){
  data_diffs[,i] <- data_diffs[,i] - data_diffs[1,i]
  data_diffs[,i] <- diff(data_df[,i])
} 
N <- nrow(data_diffs)

#==============================================================================
# Test for stationarity of first-order differences
#==============================================================================
adf.test(data_diffs$GMST)
adf.test(data_diffs$ANT)
adf.test(data_diffs$vSAOD)
adf.test(data_diffs$Nino34)
# All variables are now "significantly" stationary

# Reviewer comment: look at variation in first-order differences
df_fod <- data.frame(
  Year = rep(data_diffs$Year, 4),
  Type = factor(rep(c("GMST","Anthropogenic radiative forcing", "Volcanic SAOD", 
                      "Nino 3.4 index"), each = nrow(data_diffs)),
                levels = c("GMST","Anthropogenic radiative forcing", "Volcanic SAOD", 
                           "Nino 3.4 index")),
  Val = 100*c(data_diffs$GMST/diff(range(data_df$GMST)), 
              data_diffs$ANT/diff(range(data_df$ANT)),
              data_diffs$vSAOD/diff(range(data_df$vSAOD)),
              data_diffs$Nino34/diff(range(data_df$Nino34))) )

pdf("Figure_covariates_first_order_diff.pdf", width = 10.5, height = 4*0.9)
ggplot(df_fod, aes(x = Year, y = Val)) + 
  geom_hline(yintercept = 0, color = "gray50") + facet_grid(~Type) +
  geom_line() + theme_bw() + 
  scale_x_continuous(breaks = c(seq(from=1900,to=2000,by=25))) +
  labs(x = NULL, y = "Relative variation (%)")
dev.off()

#==============================================================================
# Analysis: no lags + first-order differences
#==============================================================================
AIC_vec <- rep(NA,15)
Pval_vec <- Pval_F <- rep(NA,15)
full_lm <- lm(GMST ~ ANT + vSAOD + Nino34, data = data_diffs)
red_lm <-  lm(GMST ~ vSAOD + Nino34, data = data_diffs)
AIC_vec[1] <- AIC(full_lm)
p <- 1
lrt <- -2*(logLik(red_lm)[1] - logLik(full_lm)[1])
Pval_vec[1] <- pchisq(lrt, df = p, lower.tail = FALSE)
SSTotal <- sum( (data_diffs$GMST - mean(data_diffs$GMST))^2 )
SSEfull     <- sum( full_lm$resid^2 )
SSRfull   <- SSTotal - SSEfull
dffull   <- full_lm$df
SSEred     <- sum( red_lm$resid^2 )
SSRred   <- SSTotal - SSEred
dfred   <- red_lm$df
Fstat <- ((SSEred - SSEfull) / (dfred - dffull)) / (SSEfull / dffull)  
Pval_F[1] <- pf(Fstat, dfred - dffull, dffull, lower.tail = FALSE)

#==============================================================================
# Analysis: with lags + first-order differences
#==============================================================================
for(p in 2:15){
  Lmat_Y <- f_create_lagmat(data_diffs$GMST, p+1)[,1:p]
  Lmat_AER <- f_create_lagmat(data_diffs$AER, p)
  Lmat_GHG <- f_create_lagmat(data_diffs$GHG, p)
  Lmat_ANT <- f_create_lagmat(data_diffs$ANT, p)
  Lmat_Volc <- f_create_lagmat(data_diffs$vSAOD, p)
  Lmat_ENSO <- f_create_lagmat(data_diffs$Nino34, p)
  Lmat_ANT <- f_create_lagmat(data_diffs$ANT, p)

  Xfull <- as.data.frame(cbind(Lmat_Y, Lmat_ANT, Lmat_Volc, Lmat_ENSO))
  Xred <- as.data.frame(cbind(Lmat_Y, Lmat_Volc, Lmat_ENSO))
  names(Xfull) <- c(paste0("Data[",1:p,"]"), paste0("ANT[",1:p,"]"), 
                    paste0("Volc[",1:p,"]"), paste0("ENSO[",1:p,"]"))
  names(Xred) <- c(paste0("Data[",1:p,"]"), paste0("Volc[",1:p,"]"), 
                   paste0("ENSO[",1:p,"]"))
  full_lm <- lm(data_diffs$GMST ~ ., data = Xfull)
  AIC_vec[p] <- AIC(full_lm)
  red_lm <- lm(data_diffs$GMST ~ ., data = Xred)

  lrt <- -2*(logLik(red_lm)[1] - logLik(full_lm)[1])
  Pval_vec[p] <- pchisq(lrt, df = p, lower.tail = FALSE)
  
  SSTotal <- sum( (data_diffs$GMST - mean(data_diffs$GMST))^2 )
  SSEfull     <- sum( full_lm$resid^2 )
  SSRfull   <- SSTotal - SSEfull
  dffull   <- full_lm$df
  SSEred     <- sum( red_lm$resid^2 )
  SSRred   <- SSTotal - SSEred
  dfred   <- red_lm$df
  Fstat <- ((SSEred - SSEfull) / (dfred - dffull)) / (SSEfull / dffull)  
  Pval_F[p] <- pf(Fstat, dfred - dffull, dffull, lower.tail = FALSE)
}

#==============================================================================
# Reconstruct using best model + Granger causality
#==============================================================================
p <- 5
Lmat_Y <- f_create_lagmat(data_diffs$GMST, p+1)[,1:p]
Lmat_AER <- f_create_lagmat(data_diffs$AER, p)
Lmat_GHG <- f_create_lagmat(data_diffs$GHG, p)
Lmat_ANT <- f_create_lagmat(data_diffs$ANT, p)
Lmat_Volc <- f_create_lagmat(data_diffs$vSAOD, p)
Lmat_ENSO <- f_create_lagmat(data_diffs$Nino34, p)

# Loop over causal variables
Pval_Final_each <- rep(NA, 3)
for(v in 1:3){
  Xfull <- as.data.frame(cbind(Lmat_Y, Lmat_ANT, Lmat_Volc, Lmat_ENSO))
  if(v == 1) Xred <- as.data.frame(cbind(Lmat_Y, Lmat_Volc, Lmat_ENSO))
  if(v == 2) Xred <- as.data.frame(cbind(Lmat_Y, Lmat_ANT, Lmat_ENSO))
  if(v == 3) Xred <- as.data.frame(cbind(Lmat_Y, Lmat_ANT, Lmat_Volc))
  full_lm <- lm(data_diffs$GMST ~ ., data = Xfull)
  red_lm <- lm(data_diffs$GMST ~ ., data = Xred)
  SSTotal <- sum( (data_diffs$GMST - mean(data_diffs$GMST))^2 )
  SSEfull     <- sum( full_lm$resid^2 )
  SSRfull   <- SSTotal - SSEfull
  dffull   <- full_lm$df
  SSEred     <- sum( red_lm$resid^2 )
  SSRred   <- SSTotal - SSEred
  dfred   <- red_lm$df
  Fstat <- ((SSEred - SSEfull) / (dfred - dffull)) / (SSEfull / dffull)  
  Pval_Final_each[v] <- pf(Fstat, dfred - dffull, dffull, lower.tail = FALSE)
}

#==============================================================================
# Regression coeffients from unrestricted model
#==============================================================================
laggedcoef_df <- data.frame(
  Group = factor(c(rep(c("Data","ANT","Volc","ENSO"), each = p)),
                 levels = c("Data","ANT","Volc","ENSO")),
  Lag = c(paste0("Lag-", (p):1), rep(paste0("Lag-", (p-1):0), 3)),
  Mean = full_lm$coefficients[-1],
  LB = full_lm$coefficients[-1] - qnorm(0.05, lower.tail = FALSE)*sqrt(diag(vcov(full_lm)))[-1],
  UB = full_lm$coefficients[-1] + qnorm(0.05, lower.tail = FALSE)*sqrt(diag(vcov(full_lm)))[-1]
)

pdf("Figure_aic_coefs.pdf", height = 4, width = 13)
grid.arrange(
  ggplot(data.frame(p=1:15,AIC=AIC_vec), aes(x = p, y = AIC)) + geom_line() + 
    geom_point() + theme_bw() + labs(x = "Lag", y = "AIC") +
    scale_x_continuous(breaks = c(1,3,5,7,9,11,13,15)) +
    ggtitle("(a) AIC scores"),
  ggplot(laggedcoef_df, aes(x = Lag, y = Mean, ymin = LB, ymax = UB)) + 
    geom_hline(yintercept = 0) + geom_pointrange() + theme_bw() +
    facet_wrap(~Group, scales = "free", nrow = 1) + xlab("") +
    ggtitle("(b) Lag-5 coefficients with 90% confidence interval"),
  nrow = 1, widths = c(1.25,4)
)
dev.off()

#==============================================================================
# Setup matrices for linear algebra
#==============================================================================
# Matrix of zeros 
Zeromat <- matrix(0, nrow=N, ncol=p)
# Matrix for cumulative sum
Jmat <- matrix(1, N, N)
for(i in 1:(N-1)) Jmat[i,(i+1):N] <- 0
# Original design matrix + inverse of X'X
Wmat <- cbind(rep(1,nrow(Lmat_Y)), Lmat_Y, Lmat_ANT, Lmat_Volc, Lmat_ENSO)
WtWinv <- solve(t(Wmat) %*% Wmat)
# Tall matrix of stacked variable-specific first-order differences
W_tall <- rbind( cbind(rep(1,nrow(Lmat_Y)),Lmat_Y, Lmat_ANT, Lmat_Volc, Lmat_ENSO), 
                 cbind(rep(0,nrow(Lmat_Y)), Lmat_Y, Zeromat, Zeromat, Zeromat),
                 cbind(rep(0,nrow(Lmat_Y)), Zeromat, Lmat_ANT, Zeromat, Zeromat),
                 cbind(rep(0,nrow(Lmat_Y)), Zeromat, Lmat_GHG, Zeromat, Zeromat),
                 cbind(rep(0,nrow(Lmat_Y)), Zeromat, Lmat_AER, Zeromat, Zeromat),
                 cbind(rep(0,nrow(Lmat_Y)), Zeromat, Zeromat, Lmat_Volc, Zeromat),
                 cbind(rep(0,nrow(Lmat_Y)), Zeromat, Zeromat, Zeromat, Lmat_ENSO) )
# Tall version of Jmat
J_tall <- diag(7) %x% Jmat
# Critical value
zalpha <- qnorm(1-1/6)

#==============================================================================
# Convert from regression coefficients back to original scale
#==============================================================================
# Best estimates
fulldata_tall <- J_tall %*% W_tall %*% full_lm$coefficients
# Standard errors
fulldata_tall_sd <- summary(full_lm)$sigma * sqrt(diag(J_tall %*% W_tall %*% WtWinv 
                                                       %*% t(W_tall) %*% t(J_tall)))

# Combine + plot
results_df <- data.frame(
  Year = rep(1901:2015, 7),
  Group = factor(rep(c("ALL","Data","ANT","GHG","AER","Volcanoes","ENSO"), each = 115),
                 levels = c("ALL","Data","ANT","GHG","AER","Volcanoes","ENSO")),
  Est = fulldata_tall, SD = fulldata_tall_sd,
  LB = fulldata_tall - zalpha*fulldata_tall_sd, 
  UB = fulldata_tall + zalpha*fulldata_tall_sd
)
for(j in c(3,5:6)){
  results_df[results_df$Group == "ENSO",j] <- results_df[results_df$Group == "ENSO",j]  
               - mean(fulldata_tall[results_df$Group == "ENSO"])
} 
results_df$Est[results_df$Group == "Data"] <- data_df$GMST[-1]
results_df$LB[results_df$Group == "Data"] <- NA # data_df$GMST[-1]
results_df$UB[results_df$Group == "Data"] <- NA # data_df$GMST[-1]

pdf("Figure_contributions.pdf", height = 6, width = 13)
grid.arrange(
  ggplot(results_df[results_df$Group == "Data",]) + 
    geom_hline(yintercept = 0, color = "gray40") + coord_cartesian(ylim=c(-0.5,1.2)) +
    geom_line(aes(x = Year, y = Est)) + labs(y = "Temperature (C)") +
    # geom_ribbon(aes(x = Year, ymin = LB, ymax = UB, fill = Group), alpha = 0.2) +
    theme_bw() + facet_wrap(~Group, nrow = 2),
  ggplot(results_df[results_df$Group != "Data",]) + 
    geom_hline(yintercept = 0, color = "gray40") + coord_cartesian(ylim=c(-0.5,1.2)) +
    geom_line(aes(x = Year, y = Est, color = Group)) +
    geom_ribbon(aes(x = Year, ymin = LB, ymax = UB, fill = Group), alpha = 0.2) +
    theme_bw() + facet_wrap(~Group, nrow = 2) + labs(y = "Temperature (C)") +
    scale_fill_manual(values = c("black", brewer.pal(5, "Set1")[c(3,1,2,4,5)]), name = "", guide = "none") +
    scale_color_manual(values = c("black", brewer.pal(5, "Set1")[c(3,1,2,4,5)]), name = "", guide = "none"),
  layout_matrix = matrix(c(1,2,NA,2), nrow=2,byrow=T), widths = c(1.075,3), heights = c(1.17,1)
)
dev.off()

#==============================================================================
# Calculate effects on the original scale
#==============================================================================
# Years for comparison (idx1 minus idx2)
idx1 <- c(115, 115, 97, 93, 1)
idx2 <- c(1,   1,   1,   1, 56)

# Setup a matrix of contrasts
cntrst <- matrix(0, N, 5)
for(k in 1:5){
  cntrst[idx1[k],k] <- 1
  cntrst[idx2[k],k] <- -1
}
cntrst_trendatt <- rbind(
  c(rep(0,N), rep(0,N), cntrst[,1], rep(0,N),   rep(0,N),   rep(0,N),   rep(0,N)), # ANT
  c(rep(0,N), rep(0,N), rep(0,N),   cntrst[,2], rep(0,N),   rep(0,N),   rep(0,N)), # GHG
  c(rep(0,N), rep(0,N), rep(0,N),   rep(0,N),   cntrst[,3], rep(0,N),   rep(0,N)), # AER
  c(rep(0,N), rep(0,N), rep(0,N),   rep(0,N),   rep(0,N),   cntrst[,4], rep(0,N)), # Volc
  c(rep(0,N), rep(0,N), rep(0,N),   rep(0,N),   rep(0,N),   rep(0,N),   cntrst[,5] )  # ENSO
)

# Calculate the effect and uncertainty
effects <- contrast_mat_trendatt %*% J_tall %*% W_tall %*% full_lm$coefficients
effects_sd <- summary(full_lm)$sigma * sqrt(diag(contrast_mat_trendatt
                                                 %*% J_tall %*% W_tall %*% WtWinv %*% t(W_tall) %*% t(J_tall)
                                                 %*% t(contrast_mat_trendatt)))

effect_analytic_df <- data.frame(
  Variable = c("ANT","GHG","AER","Volcanoes","ENSO"),
  Delta = effects, SE = effects_sd,
  LB = effects - zalpha*effects_sd,
  UB = effects + zalpha*effects_sd
)

#==============================================================================
# Event attribution: calculate scenario-specific means/SDs
#==============================================================================
cntrst_eea <- matrix(0, 4, 7*N)
cntrst_eea[1,c(2*N + 1, 5*N + 115, 6*N + 115)] <- 1    # 1901 ANT forcing, 2015 ENSO and Volcanoes
cntrst_eea[2,c(2*N + 90, 5*N + 115, 6*N + 115)] <- 1   # 1990 ANT forcing, 2015 ENSO and Volcanoes
cntrst_eea[3,c(2*N + 115, 5*N + 115, 6*N + 115)] <- 1  # 2015 ANT forcing, 2015 ENSO and Volcanoes
cntrst_eea[4,c(3*N + 115, 5*N + 115, 6*N + 115)] <- 1  # 2015 GHG forcing, 2015 ENSO and Volcanoes

# Calculate scenario means and SDs
scenario_mean <- cntrst_eea %*% fulldata_tall - mean(fulldata_tall[results_df$Group == "ENSO"])
scenario_sd <- summary(full_lm)$sigma * sqrt(diag(cntrst_eea %*% J_tall %*% W_tall %*% 
                                                    WtWinv %*% t(W_tall) %*% t(J_tall) %*% t(cntrst_eea)))

# Probability of GMST >= 0.9 (the value in 2015)
probabilities <- pnorm(0.9, mean = scenario_mean, sd = scenario_sd, lower.tail = FALSE)

# Plot
xvals <- seq(from=-1,to=4,length=500)
density_df <- data.frame(
  x = rep(xvals, 4),
  y = c(dnorm(xvals, mean = scenario_mean[1], sd = scenario_sd[1]),
        dnorm(xvals, mean = scenario_mean[2], sd = scenario_sd[2]),
        dnorm(xvals, mean = scenario_mean[3], sd = scenario_sd[3]),
        dnorm(xvals, mean = scenario_mean[4], sd = scenario_sd[4])),
  type = rep(paste0("Scenario ",1:4), each = 500)
)

pdf("Figure_eventAttribution.pdf", height = 4, width = 8)
ggplot() + 
  geom_line(data = density_df, mapping = aes(x = x, y = y, color = type)) +
  labs(x = "GMST anomaly (°C)", y = "Density") +
  scale_color_manual(values = brewer.pal(7,"Set1")[c(2,3,4,1)], name = "") + 
  scale_fill_manual(values = brewer.pal(7,"Set1")[c(2,3,4,1)], name = "") + 
  geom_vline(xintercept = 0.9, linetype = "dashed") + 
  geom_ribbon(data = density_df[density_df$x >= 0.9,], aes(fill = type, x = x, ymax = y),
              ymin = 0, alpha = 0.3) +
  annotate("label", label = c("Event of interest"), x = c(1), y = c(1.6), size = 4) +
  annotate("segment", x = c(1.5, 2, 2.5)+0.5, xend = c(0.98, 1.65, 2.3),
           y = c(1.25, 1, 0.75)-0.2, yend = c(0.15, 0.22, 0.1),
           arrow=arrow(length=unit(0.2,"cm")), size = 0.5, color = "black") +
  annotate("label", color = brewer.pal(7,"Set1")[c(2,3,4,1)], 
           label = c(expression("P"[1]~"< 0.0001"), expression("P"[2]~"= 0.067"),
                     expression("P"[3]~"= 0.46"), expression("P"[4]~"= 0.59")), 
           x = c(1.5, 2, 2.5, 3), y = c(1.5, 1.25, 1, 0.75)-0.2, size = 4) +
  theme_bw() + coord_cartesian(ylim = c(0,1.6), xlim = c(-0.5, 3.2))
dev.off()

# Bootstrap estimates for UQ
Nboot <- 500
boot_probabilities <- matrix(NA, Nboot, 4)
set.seed(0)
for(b in 1:Nboot){
  if(b %% 50 == 0) cat(b, " ")
  ind_boot <- sort(sample(nrow(data_diffs), replace = TRUE))
  boot_lm <-  lm(data_diffs$GMST[ind_boot] ~ ., data = Xfull[ind_boot,])
  boot_tall <- J_tall %*% W_tall %*% boot_lm$coefficients
  scenario_mean <- cntrst_eea %*% boot_tall - mean(boot_tall[results_df$Group == "ENSO"])
  scenario_sd <- summary(boot_lm)$sigma * sqrt(diag(cntrst_eea %*% J_tall %*% W_tall %*% WtWinv 
                                                    %*% t(W_tall) %*% t(J_tall) %*% t(cntrst_eea)))
  boot_probabilities[b,] <- pnorm(0.9, mean = scenario_mean, sd = scenario_sd, lower.tail = FALSE)
}

# Event attribution: risk ratios
RR <- c(probabilities[3]/probabilities[1], probabilities[3]/probabilities[2],
        probabilities[4]/probabilities[3])
RR_boot <- cbind(boot_probabilities[,3]/boot_probabilities[,1], 
                 boot_probabilities[,3]/boot_probabilities[,2], 
                 boot_probabilities[,4]/boot_probabilities[,3])
RR_lb <- apply(RR_boot, 2, quantile, prob = 1/6)
RR_ub <- apply(RR_boot, 2, quantile, prob = 1-1/6)

data.frame( RR = RR, LB = RR_lb, UB = RR_ub )



\end{verbatim}

\end{appendix}

\end{document}